\documentclass[aps,10pt,prd,notitlepage,nofootinbib,superscriptaddress]{revtex4-1}

\usepackage[utf8]{inputenc}
\usepackage{amsmath,amssymb,amsfonts}
\usepackage{newtxtext,newtxmath}
\usepackage{bbold}
\usepackage{bm}
\usepackage{graphicx}
\usepackage[usenames,dvipsnames]{xcolor}
\usepackage{color}
\usepackage{tikz}
\usepackage[colorlinks=true,linkcolor=blue,urlcolor=blue,citecolor=blue]{hyperref}
\usepackage{slashed}
\usepackage[english]{babel}
\usepackage{microtype}
\usepackage{dcolumn}
\usepackage{blindtext}
\usepackage{epsfig}
\usepackage{pifont}
\usepackage{dsfont,mathrsfs}
\usepackage{cancel}
\usepackage{bigints}

\usepackage{xcolor}
\usepackage{bigints}

\newcommand{\nn}{\nonumber}

\newcommand{\ket}[1]{\left|#1\right\rangle}

\newcommand{\MB}[1]{\left|#1\right|}
\newcommand{\FB}[1]{\left(#1\right)}
\newcommand{\SB}[1]{\left\{#1\right\}}
\newcommand{\TB}[1]{\left[#1\right]}

\newcommand{\mcT}{\mathcal{T}}

\newcommand{\scrL}{\mathscr{L}}
\newcommand{\scrM}{\mathscr{M}}
\newcommand{\mcV}{\mathcal{V}}

\newcommand{\munu}{{\mu\nu}}

\newcommand{\IM}{\text{Im}}
\newcommand{\RE}{\text{Re}}

\newcommand{\kpll}{k_\parallel}
\newcommand{\kper}{k_\perp}
\newcommand{\ppll}{p_\parallel}
\newcommand{\pper}{p_\perp}
\newcommand{\spll}{s_\parallel}
\newcommand{\sper}{s_\perp}
\newcommand{\Ztil}{\tilde{Z}}

\newcommand{\kperv}{|\vec{k}_\perp|}
\newcommand{\pperv}{|\vec{p}_\perp|}
\newcommand{\vac}{\text{vac}}
\newcommand{\ktilde}{\tilde{k}}
\newcommand{\ztil}{\tilde{z}}
\newcommand{\Weak}{\text{Weak}}
\newcommand{\Strong}{\text{Strong}}
\newcommand{\ktilpll}{\tilde{k}_\parallel}
\newcommand{\ktilper}{\tilde{k}_\perp}
\newcommand{\qpll}{q_\parallel}
\newcommand{\qper}{q_\perp}
\newcommand{\del}{\partial}

\begin{document}
\title{Scattering cross-section under external magnetic field using the optical theorem}

\author{Snigdha Ghosh}
\email{snigdha.physics@gmail.com, snigdha.ghosh@saha.ac.in}
\thanks{(Corresponding Author)}
\affiliation{Saha Institute of Nuclear Physics, 1/AF Bidhannagar, Kolkata - 700 064, India}
\affiliation{Indian Institute of Technology Gandhinagar, Palaj, Gandhinagar - 382 355, Gujarat, India}
\author{Vinod Chandra}
\email{vchandra@iitgn.ac.in}
\affiliation{Indian Institute of Technology Gandhinagar, Palaj, Gandhinagar - 382 355, Gujarat, India}

\begin{abstract}
The cross-section for the lowest order $2\rightarrow2$ elastic scattering between two charged scalars under external magnetic field mediated via a neutral scalar, has been computed in strong as well as weak magnetic field limits. This has been done by applying the optical theorem where the cross-section is expressed in terms of the imaginary parts of different one-loop graphs contributing to the forward scattering amplitudes. The modification in the amplitudes due to the external magnetic field has been done by means of replacing the charged scalar propagators with the Schwinger proper-time ones. Significant modifications of the cross-sections with respect to the vacuum cross-section are observed due to the external magnetic field. 
\end{abstract}

\maketitle
%
\section{Introduction}\label{sec.intro}
The dynamics and interactions among different elementary particles in the universe are governed by their respective underlying Quantum Field Theories (QFTs). The QFT in vacuum is well established and successful in describing a wide range of physical processes. However, the perfect `vacuum' is an ideal concept that is not achievable in reality. Even the whole universe is immersed in the sea of a thermal background called the Cosmic Microwave Background Radiation (CMBR). Thus, the formalism of Quantum Field Theory (QFT) in a non-trivial background has been an intense field of research over the past many decades. For example, the background, being a thermal medium, gives rise to a different branch of the QFT namely the Thermal Field Theory (TFT)~\cite{Bellac:2011kqa} which is a widely used theoretical tool for the study of a system of hot and/or dense matter. Another example is the background magnetic field, for which an exact formalism of the QFT is not well established yet specifically for the time dependent fields.

On the other hand, the study of hot nuclear matter under a strong external magnetic field has gained considerable research interest in recent years~\cite{Kharzeev:2012ph}. Such a situation might have existed in the early universe (during the electroweak phase transitions~\cite{Vachaspati:1991nm} and the Big Bang nucleosynthesis~\cite{Giovannini:2003yn,Widrow:2011hs}) and also may be relevant for the quark-matter in the core of a neutron star~\cite{Duncan:1992hi}. These studies are also of great importance because of the appearance of various novel phenomenon due to such non-trivial background such as the Chiral Magnetic Effect (CME), Chiral Vortical Effect (CVE), the Magnetic Catalysis (MC), the Inverse Magnetic Catalysis (IMC), the vacuum superconductivity, etc.~\cite{Kharzeev:2007tn,Kharzeev:2007jp, Fukushima:2008xe,Gusynin:1995nb,Gusynin:1999pq,Bali:2011qj,Chernodub:2010qx,Chernodub:2012tf}. Recent studies have revealed that the Heavy Ion Collision (HIC) experiments at Relativistic Heavy Ion Collider (RHIC) or Large Hadron Collider (LHC), have the potential to create strong magnetic fields in the laboratory~\cite{Skokov:2009qp}. For instance, the magnitude of the magnetic field could be $eB\sim 15 m_\pi^2$ in HIC at Large Hadron Collider (LHC) which is much higher that could perhaps be seen in neutron stars. Thus, it is necessary to have a proper understanding of different physical processes among the elementary particles in a background magnetic field. This sets the initial motivation for the analysis presented in this work.

The primary quantities, that one can calculate using the QFT are the decay rates and cross-sections involving the elementary particles~\cite{Peskin:1995ev}. Here, we aim to study the $2\rightarrow2$ scattering processes under the external magnetic field \textit{at zero temperature}. This could be seen as the first step in gathering necessary inputs and understanding the work to finite temperature and seek for its utilization in QED effects to hot QCD medium in the light of the plethora of interesting data from the HIC experiments. For example, the calculation of various transport coefficients (viscosities and conductivities) of the hot, dense and magnetized ``strongly" interacting matter created in HIC experiments primarily requires the knowledge of scattering cross-sections of various $2\rightarrow2$ QCD processes (since the cross-section goes as the dynamical inputs). The study and understanding of other probes of HIC like heavy quark propagation also require the estimation of scattering cross-sections in such a non-trivial background. In particular, the collisional energy loss of a heavy quark~\cite{Braaten:1991we} inside the plasma of quarks and gluons will largely depend on the cross-sections of QCD processes and the modifications of the cross-section due to the external magnetic field will certainly affect the calculated nuclear modification factor ($R_{AA}$)~\cite{Adare:2006nq,Adler:2005xv} which can be measured in the experiment.

Some of the previous works attempting to calculate the cross-section under external magnetic field are available in the literature for example, in the context of processes involving neutrino in Refs.~\cite{Arras:1998mv,Lai:1998sz,Roulet:1997sw,Bhattacharya:2002qf}, electron-muon scattering in Refs.~\cite{Binoy_Tiwari:2018uxk}, Compton scattering in Refs.~\cite{Herold:1979zz,Bussard:1986wd,Gonthier:2014wja,Dass:1975zz,Binoy_Pal:2018nxm}. It can be realized that, unlike the vacuum case where one can directly apply the Feynman rules to obtain the invariant amplitudes for the scattering processes, the same is not possible in presence of the external magnetic field. This is simply because the Feynman rules are not available/formulated for the non-zero external magnetic field. Thus, one needs to calculate the cross-section starting from the first principle using the $\hat{S}$-matrix expansion. For that, the knowledge of the Fourier decomposition of the fields in terms of creation and annihilation operators is also necessary. The decomposition has to be done on the basis of the solutions of the corresponding equation of motion under the external magnetic field (for example the Klein-Gordon equation for the scalar field or the Dirac equation for the spinor field). In other words, one has to consistently quantize the field theory in the presence of a background magnetic field. Such quantization procedures are not well established yet. In Ref.~\cite{Bhattacharya:2002qf}, the authors have calculated the cross-section for inverse beta decay using this methodology without any approximation on the strength of the magnetic field. The same procedure has also been followed in Refs.~\cite{Binoy_Tiwari:2018uxk,Binoy_Pal:2018nxm} under a strong field approximation.

In this work, we will use a different approach to calculate the cross-section which is the optical theorem. According to the optical theorem, the imaginary part of the forward scattering amplitude is proportional to the total cross-section~\cite{Peskin:1995ev}. Thus the calculation of the cross-section for $2\rightarrow2$ scattering process using the optical theorem requires the evaluation of different four-point correlation functions. The novelty of this formalism is that, it is valid even at the non-zero external magnetic field since it directly follows from the unitarity of the $\hat{S}$-matrix. Moreover, the advantage of this formalism is that, the incorporation of external magnetic field is straightforward; the propagators for the charged particles have to be replaced by the respective magnetized propagators (for example, the Schwinger proper-time propagator~\cite{Schwinger:1951nm}). The same methodology also applies to the calculation of decay widths in the presence of an external magnetic field. In that case, one obtains the one-loop self-energy of the parent particle under external magneic field. The imaginary part of the self-energy leads to the decay width~\cite{Aguirre:2018fbo,Bandyopadhyay:2016cpf, Bandyopadhyay:2018gbw,Hattori:2012je,Kawaguchi:2016gbf,Piccinelli:2017yvl,Jaber-Urquiza:2018oex, Ghosh:2016evc,Ghosh:2017rjo,Ghosh:2019fet}.

Although, the study of cross-sections for the different elementary processes in Quantum Electrodynamics (QED) or Quantum Chromodynamics (QCD) would be more relevant and interesting, yet in this work, we have considered the scattering of two charged scalar bosons ($b^+b^-\rightarrow b^+b^-$) mediated by a neutral scalar boson ($B^0$). This particular process has been chosen to avoid the additional complications arising from the spins of the particle. Nevertheless, it will guide us for the future generalizations to tackle the hot QCD matter in the influence of external electromagnetic fields. We have first obtained the cross-section using the optical theorem by explicitly calculating the imaginary parts of the forward scattering amplitudes at \textit{zero external magnetic field}. For a consistency check, the same has been calculated from the invariant amplitudes of the tree-level diagrams. Both methods lead to an identical cross-section. At \textit{non-zero the external magnetic field}, the propagators for the charged scalars are replaced with the Schwinger proper-time ones. The calculations under the external magnetic field are done for the weak field and the strong field approximations separately. We have shown that the scattering cross-section has non-trivial dependence on the external magnetic field.

The paper is organized as follows. In Sec.~\ref{sec.tree}, the vacuum cross-section is obtained by evaluating the invariant amplitudes of the tree-level diagrams. Next, in Sec.~\ref{sec.loop}, the same has been calculated using the optical theorem by means of evaluating the forward scattering amplitudes at zero magnetic field. Then, in Sec.~\ref{sec.loop.eb}, we have introduced the external magnetic field through the Schwinger proper-time formalism and the forward scattering amplitudes are calculated in presence of external magnetic field employing both the weak and strong field approximations. This is followed by Sec.~\ref{sec.results} where we presented the numerical results for our analysis. Finally we summarize and conclude in Sec.~\ref{sec.summary}. Some computational details are provided in the appendices.

\section{VACUUM CROSS-SECTION FROM THE TREE GRAPHS} \label{sec.tree}
Let us consider the interaction of a real scalar field $\Phi(x)$ with the two complex scalar fields $\phi(x)$ and $\phi^\dagger(x)$ via the interaction Lagrangian (density)
\begin{eqnarray}
\scrL_\text{int} = g\Phi\phi\phi^\dagger \label{eq.Lagrangian}
\end{eqnarray}
where $g$ is the coupling constant which gives the strength of the above interaction. The field $\Phi$ annihilates and creates the neutral scalar particle $B^0$ whereas the field $\phi$ $(\phi^\dagger)$ annihilates (creates) the charged scalar particle $b^+$ and creates (annihilates) its antiparticle $b^-$. Few comments on the choice and applicability of the above Lagrangian are in order here. The Lagrangian in Eq.~\eqref{eq.Lagrangian} involving the simplest trilinear coupling among scalar fields corresponds to the description of a toy model. However, such trilinear interactions involving fermions are often useful; for example in Yukawa theory and other QCD motivated models like Nambu--Jona-Lasinio (NJL) model~\cite{Nambu:1961fr,Nambu:1961tp,Klevansky:1992qe}, Quark Meson (QM) model~\cite{Rabhi:2011mj,Andersen:2012bq,Fraga:2008qn}, Linear Sigma Model (LSM)~\cite{GellMann:1960np,Lenaghan:2000ey} and so on. This Lagrangian may also be useful for the effective models describing hadronic interactions~\cite{Krehl:1999km} among different mesons (like $\pi, \sigma, K, \rho$ mesons, etc.) and baryons. Though a Yukawa type Lagrangian instead of the simplest one in Eq.~\eqref{eq.Lagrangian} would have been of more physical importance, we have chosen the later solely to avoid complications arising from the spins of the particles.

\begin{figure}[h]
	\begin{center}
		\includegraphics[angle=0, scale=0.8]{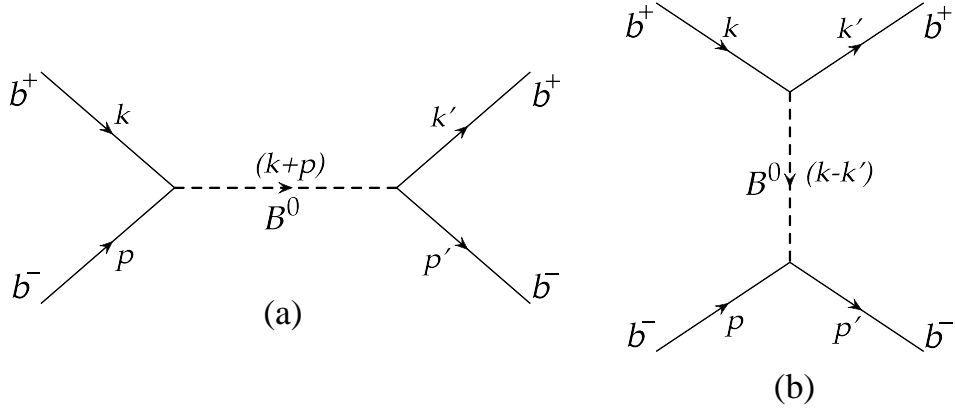}
	\end{center}
	\caption{Tree-level Feynman diagrams for the scattering process $b^+(k)b^-(p)\rightarrow b^+(k')b^-(p')$ in the (a) $s$-channel and (b) $t$-channel.}
	\label{fig.Tree}
\end{figure}

We are interested in calculating the cross-section for the elastic scattering $b^+(k)b^-(p)\rightarrow b^+(k')b^-(p')$. We restrict ourselves to the lowest order $\mathcal{O}(g^4)$ contribution. Fig.~\ref{fig.Tree} shows the corresponding tree-level Feynman diagrams for this process.
Fig.~\ref{fig.Tree}(a) and (b) respectively represents the $s$-channel and $t$-channel diagrams. Applying Feynman rules, we get the invariant amplitudes for the two channels as
\begin{eqnarray}
\scrM_s = g^2\TB{\frac{1}{s-M^2+i\epsilon}} ~\text{and} ~~
\scrM_t = g^2\TB{\frac{1}{t-M^2+i\epsilon}}
\end{eqnarray}
respectively, where $M$ is the mass of $B^0$ and the Mandelstam variables are defined as
\begin{eqnarray}
s=(k+p)^2=(k'+p')^2, \\
t=(k-k')^2=(p-p')^2, \\
u=(k-p')^2=(p-k')^2 
\end{eqnarray}
which satisfy the constraint $(s+t+u)=4m^2$ with $m$ being the mass of $b^\pm$. 
The metric tensor that has been used has the signature, $g^\munu=\text{diag}(1,-1,-1,-1)$. 
The total cross-section $\sigma(s)$ is immediately 
obtained from the invariant amplitude using the relation~\cite{Peskin:1995ev}
\begin{eqnarray}
\sigma_\text{Tree}(s) = \frac{1}{16\pi\lambda(s,m^2,m^2)}\int_{-\lambda(s,m^2,m^2)/s}^{0}dt \MB{\scrM_s+\scrM_t}^2
\label{eq.croos.tree}
\end{eqnarray}
where, $\lambda(x,y,x)=(x^2+y^2+z^2-2xy-2yz-2zx)$ is the K\"all\'en function.


\section{VACUUM CROSS-SECTION FROM THE LOOP GRAPHS USING THE OPTICAL THEOREM} \label{sec.loop} 
\begin{figure}[h]
	\begin{center}
		\includegraphics[angle=0, scale=0.5]{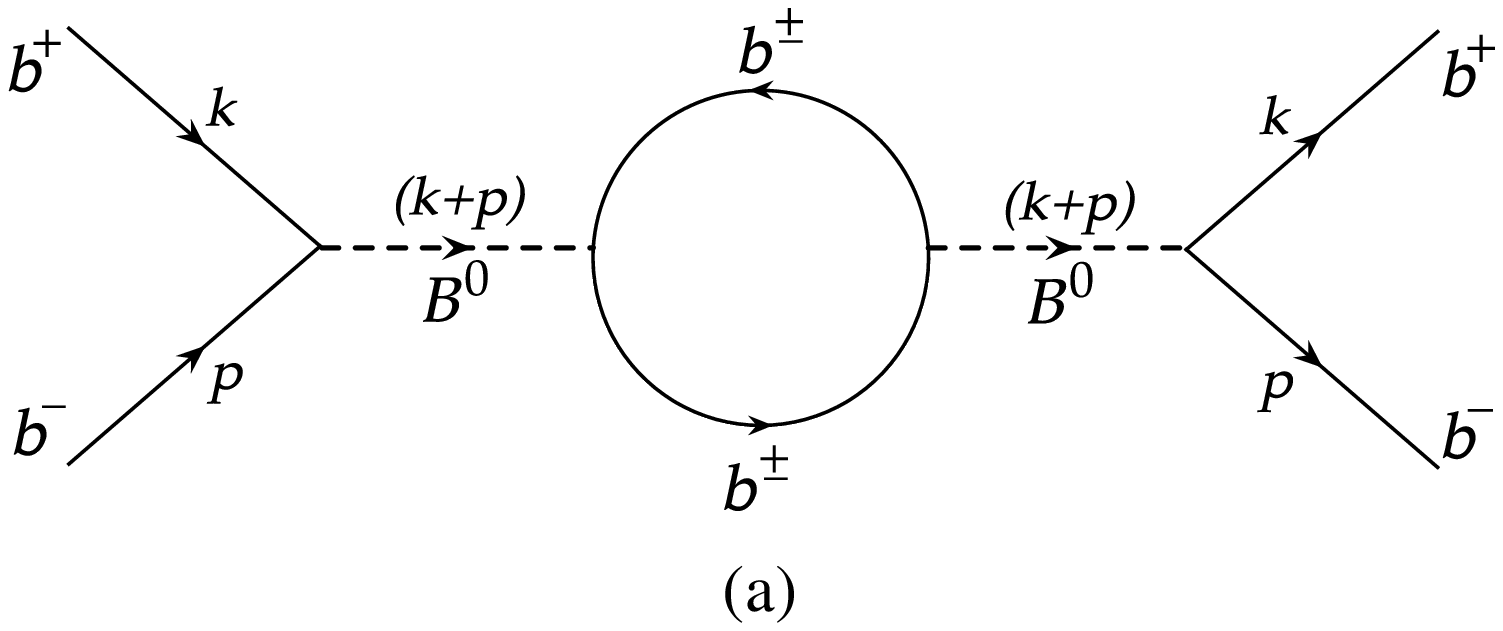} ~~~~~~~~~\includegraphics[angle=0, scale=0.4]{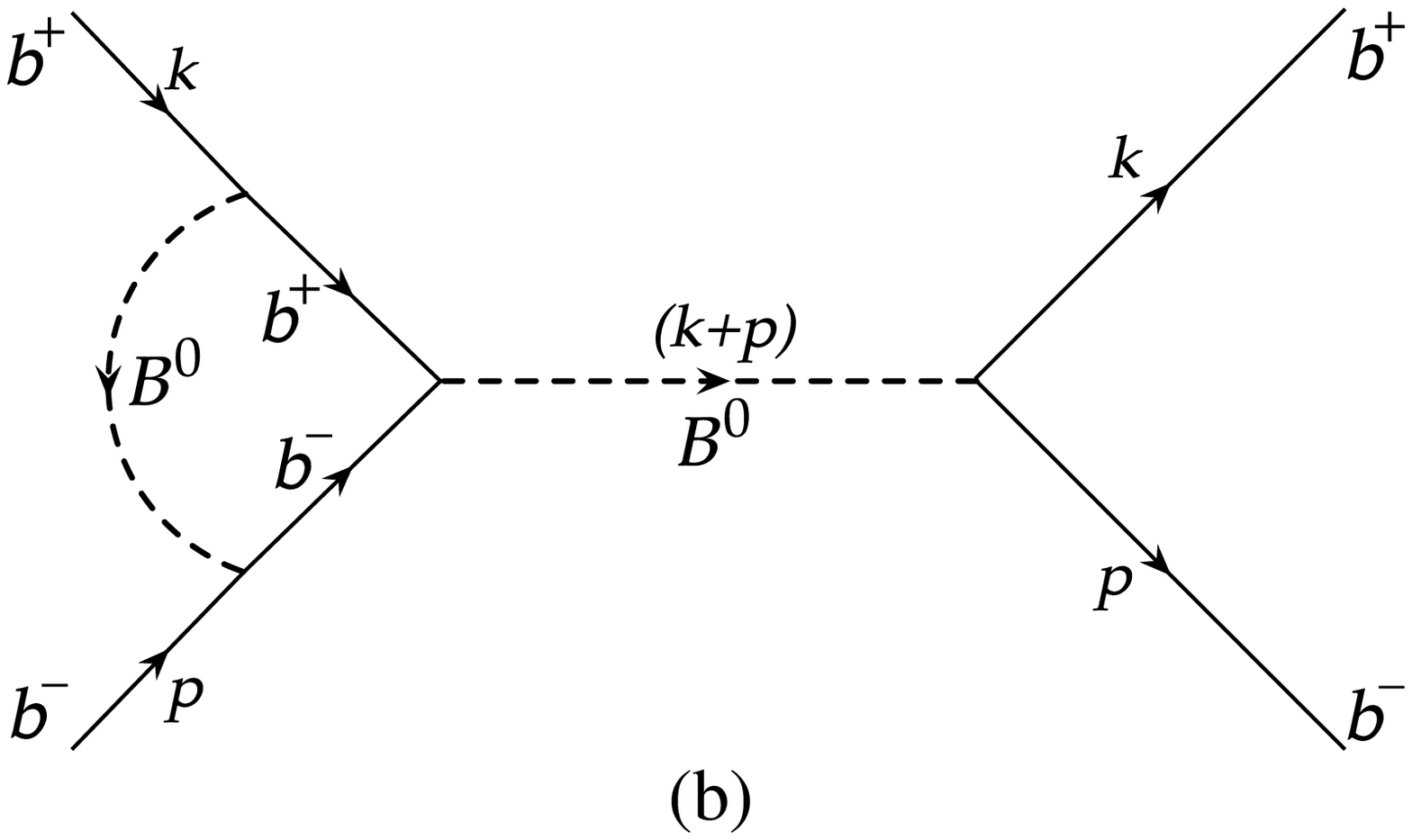} \\
		\includegraphics[angle=0, scale=0.4]{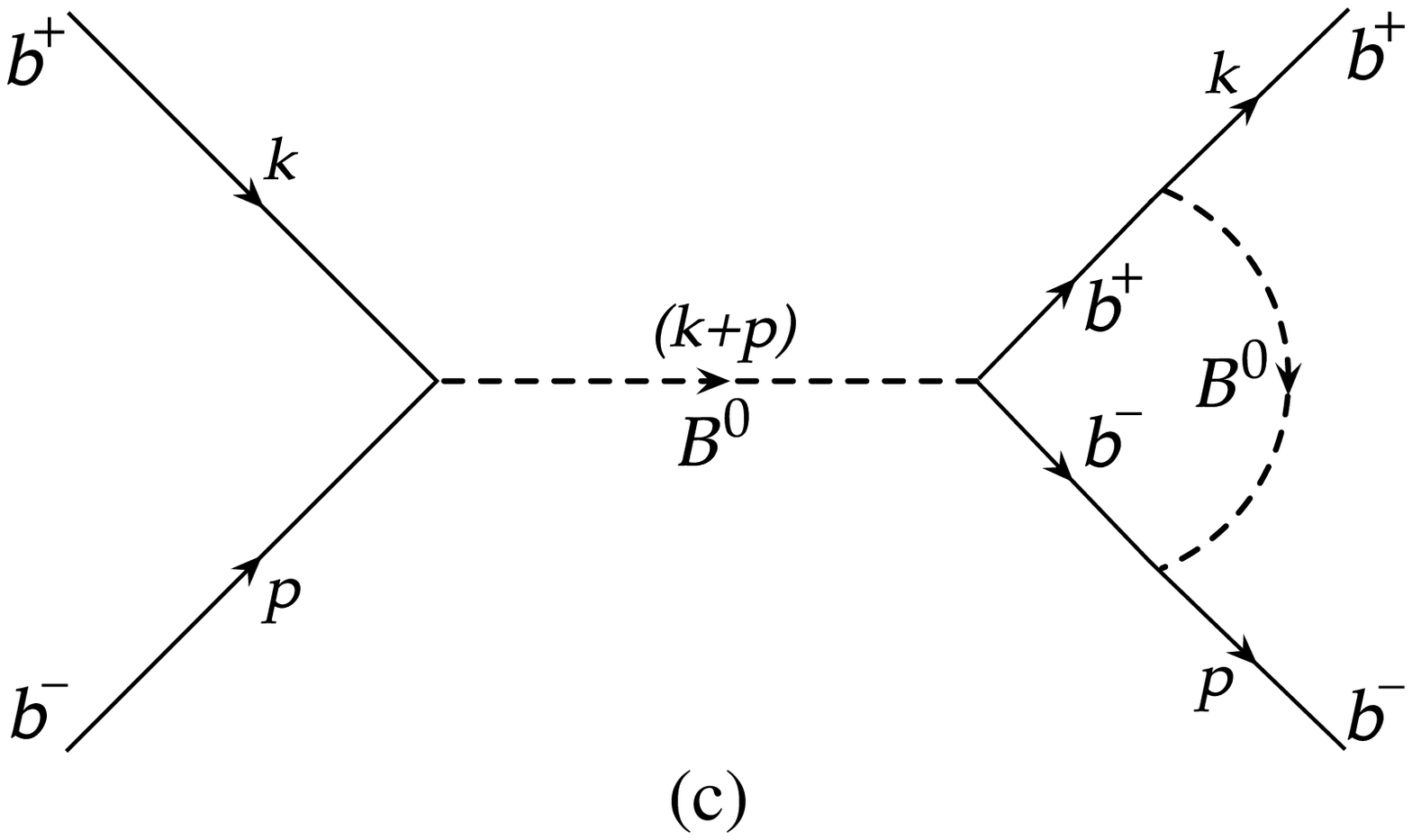} ~~~~~~~~~\includegraphics[angle=0, scale=0.4]{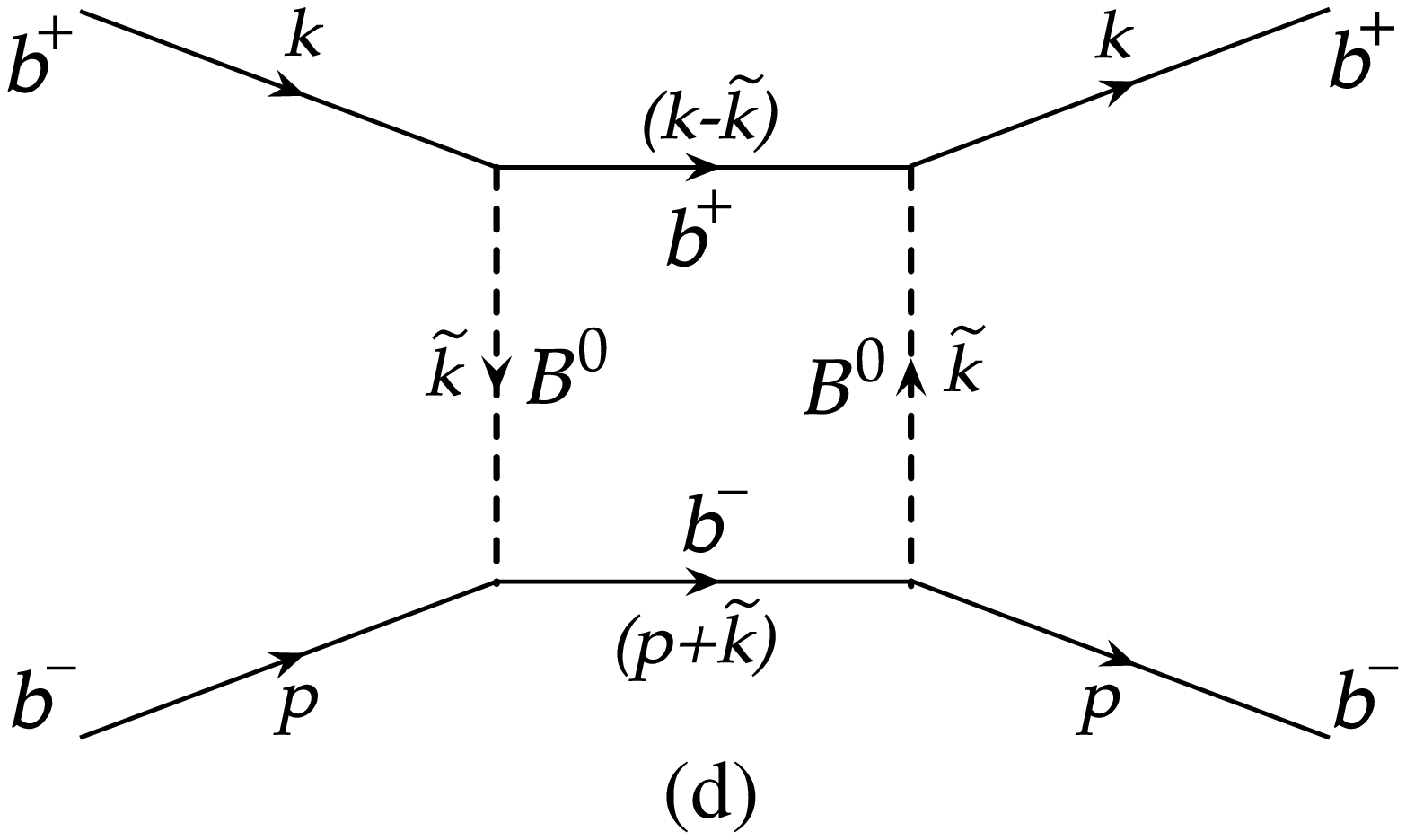} 		
	\end{center}
	\caption{One-loop Feynman diagrams contributing to forward scattering amplitude of the optical theorem corresponding to the scattering process $b^+(k)b^-(p)\rightarrow b^+(k')b^-(p)$. Subfigure (a) corresponds to the pure $s$-channel, (b) and (c) correspond to the interference of $s$-channel and $t$-channel and (d) corresponds to the 
		 pure $t$-channel.}
	\label{fig.Loop}
\end{figure}

\begin{figure}[h]
	\begin{center}
		\includegraphics[angle=0, scale=0.8]{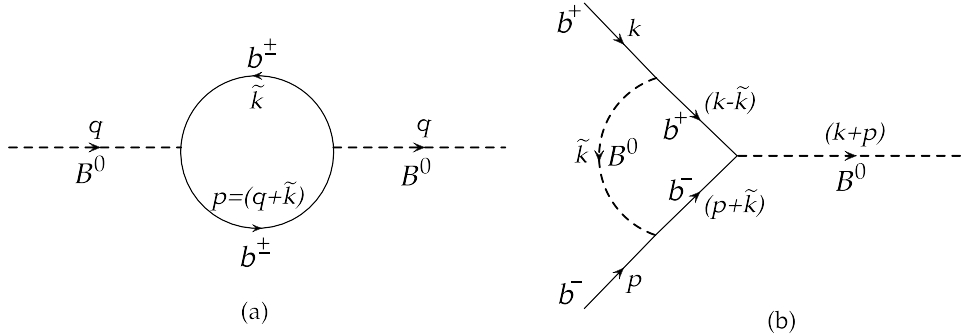}
	\end{center}
	\caption{Feynman diagram for the one-loop (a) self-energy of $B^0$ due to $b^+b^-$ loop and (b) $B^0b^+b^-$ vertex function.}
	\label{fig.SelfEnergyVertex}
\end{figure}

A possible alternative to calculate the cross-section is the optical theorem~\cite{Peskin:1995ev}, according to which, the imaginary part of the forward scattering amplitude is proportional to the total cross-section. Fig.~\ref{fig.Loop} shows the one-loop Feynman diagrams in the lowest order $\mathcal{O}(g^4)$ which contributes to the scattering process $b^+b^-\rightarrow b^+b^-$. Figs.~\ref{fig.Loop}(a) and (d) correspond to the pure $s$-channel and pure $t$-channel contributions respectively whereas Figs.~\ref{fig.Loop}(b) and (c) correspond to the interference of $s$-channel and $t$-channel.
We denote the amplitude for the diagrams given in Figs.~\ref{fig.Loop}(a), (b), (c) and (d) by $\Pi_A$, $\Pi_B$, $\Pi_C$ and $\Pi_D$ respectively. It can be noticed that, the diagrams in Figs.~\ref{fig.Loop}(a)-(c) are not 1PI (One Particle Irreducible) whereas (d) is. Applying Feynman rule, we get their respective amplitudes in vacuum as,
\begin{eqnarray}
\Pi_A^\vac(k,p) &=& g^2\TB{\frac{1}{s-M^2+i\epsilon}}^2\Pi_\vac(k+p)~, \label{eq.A.vac}\\
\Pi_B^\vac(k,p) &=& -g\TB{\frac{1}{s-M^2+i\epsilon}}\mcV_\vac(k,p)~, \label{eq.B.vac} \\
\Pi_C^\vac(k,p) &=& -g\TB{\frac{1}{s-M^2+i\epsilon}}\mcV_\vac(-k,-p) = \Pi_B^\vac(-k,-p)~, \label{eq.C.vac} \\
\Pi_D^\vac(k,p) &=& ig^4\!\!\int\!\!\frac{d^4\ktilde}{(2\pi)^4}\Delta^2_F(\ktilde,M)\Delta_F(k-\ktilde,m)\Delta_F(p+\ktilde,m) \label{eq.D.vac}
\end{eqnarray}
where, $\Delta_F(k,m)=\FB{\frac{-1}{k^2-m^2+i\epsilon}}$ is the vacuum scalar Feynman propagator; $\Pi_\vac(q)$ and $\mcV_\vac(k,p)$ are respectively the one-loop vacuum self-energy of $B^0$ and one-loop $B^0b^+b^-$ vacuum vertex function. The Feynman diagrams contributing to the one-loop $B^0$ self-energy and one-loop $B^0b^+b^-$ vertex function are shown in Fig.~\ref{fig.SelfEnergyVertex} from which we get (applying Feynman rules) 
\begin{eqnarray}
\Pi_\vac(q) &=& ig^2\!\!\int\!\!\frac{d^4\ktilde}{(2\pi)^4}\Delta_F(\ktilde,m)\Delta_F(q+\ktilde,m)~, \label{eq.Pi.Vac}\\
\mcV_\vac(k,p) &=& ig^3\!\!\int\!\!\frac{d^4\ktilde}{(2\pi)^4}\Delta_F(\ktilde,M)\Delta_F(k-\ktilde,m)\Delta_F(p+\ktilde,m)~. \label{eq.Vertex.Vac}
\end{eqnarray}

The total cross-section $\sigma(s)$ is obtained from the optical theorem as~\cite{Peskin:1995ev}
\begin{eqnarray}
\sigma_\text{Optical}(s) = \frac{-1}{\lambda^{1/2}(s,m^2,m^2)}\IM\FB{\Pi_A^\vac+\Pi_B^\vac+\Pi_C^\vac+\Pi_D^\vac}~.
\label{eq.optical}
\end{eqnarray}
Since we are only interested in calculating the cross-section, we only need to evaluate the imaginary parts of the amplitudes given in Eqs.~\eqref{eq.A.vac}-\eqref{eq.D.vac}. The calculations of imaginary parts of these forward scattering amplitudes in the vacuum \textit{i.e. in the absence of external magnetic field} is provided in Appendix~\ref{appendix.impiABCD.vac} and the final results can be read off from Eqs.~\eqref{eq.ImA.Final}, \eqref{eq.ImB.Final}, \eqref{eq.ImC.Final} and \eqref{eq.ImD.Final} as
\begin{eqnarray}
\IM\Pi_A^\vac(s) &=& -\frac{g^4}{16\pi s(s-M^2)^2}\sqrt{s-4m^2}\Theta(s-4m^2)~, \\
\IM\Pi_B^\vac(s) &=& \IM\Pi_C^\vac(s) = \frac{g^4}{16\pi(s-M^2)}\int_{0}^{1}\!\!dy\frac{\Theta(z_+)\Theta(1-y-z_+)+\Theta(z_-)\Theta(1-y-z_-)}
{\sqrt{(M^2+sy)^2-4m^2(M^2+sy^2)}}~, \\
\Pi_D^\vac(s) &=& \frac{-g^4}{16\pi}\int_{0}^{1}\!\!dy\frac{(sy+M^2-2m^2)\TB{\Theta(z_+)\Theta(1-y-z_+) 
+ \Theta(z_-)\Theta(1-y-z_-)}}{\TB{(M^2+sy)^2-4m^2(M^2+sy^2+\lambda)}^{3/2}} 
\end{eqnarray}
where, $z_\pm(s,y) = \frac{1}{2m^2}\TB{M^2-2m^2y+sy\pm\sqrt{(M^2+sy)^2-4m^2(M^2+sy^2)}}$.

It can be noticed that, the $\IM\Pi_{A,B,C,D}^\vac(k,p)$ depends only on the Lorentz scalar $s=(k+p)^2$. It is now straightforward to obtain the cross-section using the optical theorem as given in Eq.~\eqref{eq.optical}. It is expected that the cross-section obtained from the optical theorem ($\sigma_\text{Optical}(s)$) will agree with that of obtained from the tree-level calculations in Eq.~\eqref{eq.croos.tree}. We will discuss this in Sec.~\ref{sec.results}.

\section{CROSS-SECTION UNDER EXTERNAL MAGNETIC FIELD FROM THE OPTICAL THEOREM} \label{sec.loop.eb}
Now we proceed to obtain the cross-section under an external magnetic field. In this case, the calculation of invariant amplitudes from the tree graphs (as shown in Fig.~\ref{fig.Tree}) is not possible since the Feynman rules are not available for the non-zero external magnetic field. Thus we need to calculate the $\hat{S}$-matrix element $\langle f|\hat{S}|i\rangle$ from first principle for the transition from the initial state $\ket{i}=\ket{b^+(k),b^-(p)}$ to the final state $\ket{f}=\ket{b^+(k'),b^-(p')}$. The formalism for such a calculation of cross-section from $\langle f|\hat{S}|i\rangle$ has not been well established yet. Also, the calculation of $\hat{S}$-matrix element would require the knowledge of the solutions of Klein-Gordan equation under the external magnetic field. In Ref.~\cite{Bhattacharya:2002qf,Binoy_Pal:2018nxm,Binoy_Tiwari:2018uxk}, the authors have tried to proceed in this direction. However, in this work, we aim to calculate the cross-section under the external magnetic field using the optical theorem. The optical theorem is widely used for the calculation of decay widths in the presence of external magnetic field. In that case, one obtains the one-loop self-energy of the parent particle under external magneic field. The imaginary part of the self-energy leads to the decay width
~\cite{Aguirre:2018fbo,Bandyopadhyay:2016cpf, Bandyopadhyay:2018gbw,Hattori:2012je,Kawaguchi:2016gbf,Piccinelli:2017yvl,Jaber-Urquiza:2018oex, Ghosh:2016evc,Ghosh:2017rjo,Ghosh:2019fet}.

\begin{figure}[h]
	\begin{center}
		\includegraphics[angle=0, scale=0.95]{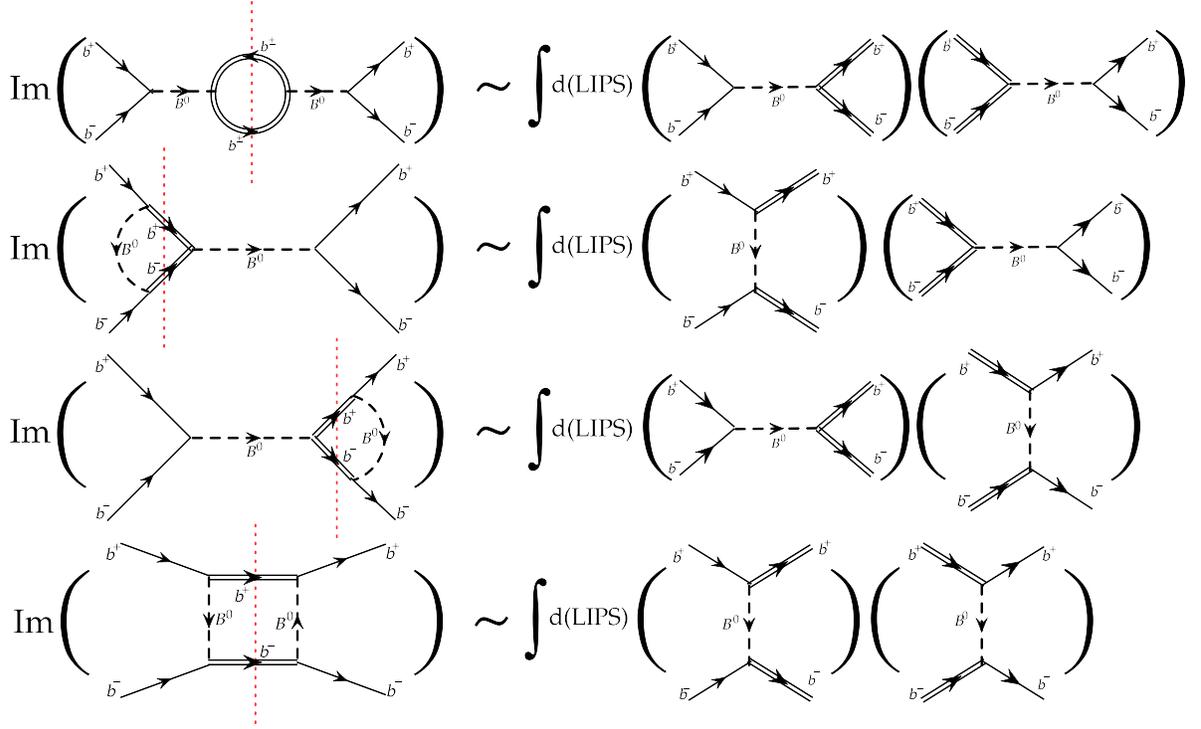}
	\end{center}
	\caption{The Cutkosky rule is used to relate the imaginary parts of the forward scattering amplitudes to the various tree-level processes. The double lines indicate the magnetic field modified propagator of charged scalar (Schwinger propagator). Here, $\int$d(LIPS) symbolically corresponds to the integration over the Lorentz invariant phase space.}
	\label{fig.ABCD}
\end{figure}

Following the same strategy, in this case, the charged scalar Feynman propagator in the loop graphs will be replaced by the corresponding propagator under the external magnetic field and thus a specific knowledge of the solutions of Klein-Gordan equation under external magnetic field is not required. 
Moreover, to ensure that this procedure indeed takes into account the magnetic modification to the external legs of the charged scalars in the tree-level diagrams of Fig.~\ref{fig.Tree}, we resort to the Cutkosky rules~\cite{Peskin:1995ev} as shown in Fig.~\ref{fig.ABCD}. From Fig.~\ref{fig.ABCD}, it is clear that the magnetic modification of the charged scalar propagators in the forward scattering graphs (denoted by the double lines) actually correspond to the modification of the external legs on the tree-level diagrams.

The propagation of charged scalar particle under external magnetic field is well described using the Schwinger proper-time formalism~\cite{Schwinger:1951nm}. The momentum space Schwinger propagator $\Delta_B$ for a charged scalar is given by a sum over infinite Landau levels~\cite{Ayala:2004dx}:
\begin{eqnarray}
\Delta_B(k,m) = -\sum_{l=0}^{\infty}\frac{-2(-1)^le^{-\alpha_k}L_l(2\alpha_k)}{\kpll^2-m^2-(2l+1)eB+i\epsilon}
\label{eq.Schwinger}
\end{eqnarray}
where $\alpha_k=-\kper^2/eB$ and $L(z)$ represents the Laguerre polynomial. Here we consider the external magnetic field $\vec{B}=B\hat{z}$ to be along the positive z-direction so that any four vector $a^\mu\equiv(a^0,a^1,a^2,a^3)$ is decomposed into $a=(a_\parallel+a_\perp)$ with $a_\parallel^\mu\equiv(a^0,0,0,a^3)$ and $a_\perp^\mu\equiv(0,a^1,a^2,0)$. The metric tensor is also decomposed into $g^\munu=(g_\parallel^\munu+g_\perp^\munu)$ where $g_\parallel^\munu=\text{diag}(1,0,0,-1)$ and $g_\perp^\munu=\text{diag}(0,-1,-1,0)$ so that $a_\parallel^\mu=g_\parallel^\munu a_\nu$ and $a_\perp^\mu=g_\perp^\munu a_\nu$. 
This makes $a_\parallel\cdot a_\parallel = a_\parallel^2=(a^0)^2-(a^3)^2$, $a_\perp\cdot a_\perp = a_\perp^2=-(a^1)^2-(a^2)^2$ and $a\cdot a = a^2=(a_\parallel^2 + a_\perp^2)$. 
It should be noted that, in our convention, $a_\perp^2$ carries an extra minus sign when compared with the convention used in Ref.~\cite{Ayala:2004dx}.
The propagator in Eq.~\eqref{eq.Schwinger} is valid for any arbitrary values of external magnetic field. However, in this work we restrict ourselves to two possible approximations namely: (a) the weak and (b) the strong field approximation. In the strong field approximation ($eB \gg m^2$), we consider only the contributions from the Lowest Landau Level (LLL), for which the propagator is
\begin{eqnarray}
\Delta_\text{Strong}(\kpll,\kper,m) = \frac{-2e^{-\alpha_k}}{\kpll^2-m^2-eB+i\epsilon}
\label{eq.Schwinger.LLL}
\end{eqnarray}
whereas, for the weak field approximation ($eB \ll m^2$), we will use the following weak field expansion~\cite{Ayala:2004dx} of the propagator (upto $\mathcal{O}(B^2)$)
\begin{eqnarray}
\Delta_\text{Weak}(\kpll,\kper,m) = \Delta_F(k,m)+(eB)^2\frac{(\kpll^2-\kper^2-m^2)}{(k^2-m^2+i\epsilon)^4}~.
\label{eq.Schwinger.Weak}
\end{eqnarray}

We now proceed to calculate the amplitudes as given in Eqs.~\eqref{eq.A.vac}-\eqref{eq.D.vac} under external magnetic field. Under the Strong/Weak field approximation, they become (by replacing the charged scalar propagators with the Schwinger one)
\begin{eqnarray}
\Pi_A^\text{Strong/Weak}(k,p) &=& g^2\TB{\frac{1}{s-M^2+i\epsilon}}^2\Pi_\text{Strong/Weak}(k+p)~, \label{eq.A.eB}\\
\Pi_B^\text{Strong/Weak}(k,p) &=& -g\TB{\frac{1}{s-M^2+i\epsilon}}\mcV_\text{Strong/Weak}(k,p)~, \label{eq.B.eB} \\
\Pi_C^\text{Strong/Weak}(k,p) &=& -g\TB{\frac{1}{s-M^2+i\epsilon}}\mcV_\text{Strong/Weak}(-k,-p) = \Pi_B^\text{Strong/Weak}(-k,-p)~, \label{eq.C.eB} \\
\Pi_D^\text{Strong/Weak}(k,p) &=& ig^4\!\!\int\!\!\frac{d^4\ktilde}{(2\pi)^4}\Delta^2_F(\ktilde,M)\Delta_\text{Strong/Weak}(k-\ktilde,m)\Delta_\text{Strong/Weak}(p+\ktilde,m) \label{eq.D.eB}
\end{eqnarray}
where, $\Pi_\text{Strong/Weak}(q)$ and $\mcV_\text{Strong/Weak}(k,p)$ are respectively the one-loop self-energy of $B^0$ and one-loop $B^0b^+b^-$ vertex function in Strong/Weak field approximation. They are obtained by replacing the charged scalar propagators in Eqs.~\eqref{eq.Pi.Vac} and \eqref{eq.Vertex.Vac} with the Schwinger propagator as
\begin{eqnarray}
\Pi_\text{Strong/Weak}(q) &=& ig^2\!\!\int\!\!\frac{d^4\ktilde}{(2\pi)^4}\Delta_\text{Strong/Weak}(\ktilde,m)\Delta_\text{Strong/Weak}(q+\ktilde,m)~, \label{eq.Pi.eB}\\
\mcV_\text{Strong/Weak}(k,p) &=& ig^3\!\!\int\!\!\frac{d^4\ktilde}{(2\pi)^4}\Delta_F(\ktilde,M)\Delta_\text{Strong/Weak}(k-\ktilde,m)\Delta_\text{Strong/Weak}(p+\ktilde,m)~. \label{eq.Vertex.eB}
\end{eqnarray}

The total cross-section is then followed from the optical theorem as
\begin{eqnarray}
\sigma_\text{Strong/Weak} = \frac{-1}{\lambda^{1/2}(s,m^2,m^2)}\IM\FB{\Pi_A^\text{Strong/Weak}+\Pi_B^\text{Strong/Weak}
	+\Pi_C^\text{Strong/Weak}+\Pi_D^\text{Strong/Weak}}~.
\label{eq.optical.eB}
\end{eqnarray}
As can be seen from Eq.~\eqref{eq.optical}, at $B=0$, the total cross-section depends only on the Lorentz scalar $s=(k+p)^2$. It can be understood in the following way: Out of the two four-vectors $k^\mu$ and $p^\mu$, the three possible Lorentz scalars that could be formed are $k^2$, $p^2$ and $k\cdot p$. However, the on-shell conditions $k^2=p^2=m^2$ require the existence of only one independent Lorentz scalar. One may choose the $s=(k+p)^2$ as the independent one. In the case of $B\ne0$, the available four vectors are $\kpll^\mu$, $\kper^\mu$, $\ppll^\mu$ and $\pper^\mu$. Correspondingly, we can construct six possible Lorentz scalars as: $\kpll^2$, $\kper^2$, $\ppll^2$, $\pper^2$, $\kpll\cdot\ppll$ and $\kper\cdot\pper$. However, the on-shell conditions $(\kpll^2+\kper^2)=(\ppll^2+\pper^2)=m^2$ require the existence of only four independent Lorentz scalars. We choose the following four as the independent ones
\begin{eqnarray}
\spll = (\kpll+\ppll)^2,~~ \sper =(\kper+\pper)^2,~~ \kper^2 ~~ \text{and}~~ \pper^2~. 
\label{eq.Mandelstam.eB}
\end{eqnarray}
Thus the total cross-section in presence of external magnetic field will be function of 
\begin{equation}
\sigma_\text{Strong/Weak}=\sigma_\text{Strong/Weak}(\spll,\sper,\kper^2,\pper^2)~.
\label{eq.Optical.Theorem.eB}
\end{equation}
Since we will be calculating the cross-sections, we only need to evaluate the imaginary parts of the different amplitudes given in Eqs.~\eqref{eq.A.eB}-\eqref{eq.D.eB}. The calculations of imaginary parts of these forward scattering amplitudes under the external magnetic field employing both the weak as well as strong field approximation are provided in Appendices~\ref{appendix.impiABCD.weak} and \ref{appendix.impiABCD.strong}.

In the weak field approximation, we have from Eqs.~\eqref{eq.A.Weak}, \eqref{eq.B.Weak}, \eqref{eq.C.Weak} and \eqref{eq.D.Weak}:
\begin{eqnarray}
\IM\Pi_A^\Weak(\spll,\sper) &=& \IM\Pi_A^\vac(s) +
\frac{g^4}{(s-M^2)^2}\frac{(eB)^2}{24\pi}\frac{\TB{s(s-4m^2)(\sper-\spll)+12m^4\sper}\Theta(s-4m^2)}{[s(s-4m^2)]^{5/2}}~, \label{eq.impiA.weak}\\
\IM\Pi_B^\Weak(\spll,\sper,\kper^2,\pper^2)&=& \IM\Pi_C^\Weak(\spll,\sper,\kper^2,\pper^2) \nn \\
&=& \IM\Pi_B^\vac(s)+\frac{g^4}{(s-M^2)}\frac{(eB)^2}{48\pi}\!\int_{0}^{1}\!\!dy
\TB{\Theta(z_+)\Theta(1-y-z_+)+\Theta(z_-)\Theta(1-y-z_-)} \nn \\ && \hspace{7cm} \times
\mathcal{T}_B^\Weak(\spll,\sper,\kper^2,\pper^2)~, \label{eq.impiBC.weak} \\
\IM\Pi_D^\Weak(\spll,\sper,\kper^2,\pper^2) &=& \IM\Pi_D^\vac(s) + g^4\frac{5(eB)^2}{28\pi}
\!\int_{0}^{1}\!\!dy\TB{\Theta(z_+)\Theta(1-y-z_+)+\Theta(z_-)\Theta(1-y-z_-)} \nn \\ && \hspace{7cm} \times
\mathcal{T}_D^\Weak(\spll,\sper,\kper^2,\pper^2) \label{eq.impiD.weak}
\end{eqnarray}
where, $z_\pm(s,y) = \frac{1}{2m^2}\TB{M^2-2m^2y+sy\pm\sqrt{(M^2+sy)^2-4m^2(M^2+sy^2)}}$;
$\mathcal{T}_B^\Weak$ and $\mathcal{T}_D^\Weak$ can be read off from Eqs.~\eqref{eq.TB.Weak} and \eqref{eq.TD.Weak}.

On the other hand, in the strong field approximation, the final results can be read off from Eqs.~\eqref{eq.A.Strong}, \eqref{eq.B.Strong}, \eqref{eq.C.Strong} and \eqref{eq.D.Strong} as
\begin{eqnarray}
\IM\Pi_A^\Strong(\spll,\sper) &=& 
\frac{-g^4}{(s-M^2)^2}\frac{eB}{4\pi}\exp\FB{\frac{\sper}{2eB}}\frac{\Theta(\spll-4m^2-4eB)}{\sqrt{\spll(\spll-4m^2-4eB)}}~, \label{eq.impiA.strong}\\
\IM\Pi_B^\Strong(\spll,\sper,\kper^2,\pper^2) &=& \IM\Pi_C^\Strong(\spll,\sper,\kper^2,\pper^2) \nn \\
&=& -\frac{g^4}{(s-M^2)}\frac{eB}{8\pi}\exp\TB{\frac{\kper^2+\pper^2}{eB}}
\int_{0}^{\infty}\!\!d\xi e^{-\xi}I_0\FB{\sqrt{\frac{\xi}{2eB}\SB{\frac{}{}\sper-2\kper^2-2\pper^2}}} 
\nn \\ && \hspace{0cm} \times
\int_{0}^{1}\!\!dy\TB{\Theta(Z_+)\Theta(1-y-Z_+)+\Theta(Z_-)\Theta(1-y-Z_-)}\mathcal{T}_B^\Strong(\spll,\kper^2,\pper^2) ~,
\label{eq.impiBC.strong} \\
\IM\Pi_D^\Strong(\spll,\sper,\kper^2,\pper^2) &=& -g^4\frac{eB}{8\pi}\exp\TB{\frac{\kper^2+\pper^2}{eB}}
\int_{0}^{\infty}\!\!d\xi e^{-\xi}I_0\FB{\sqrt{\frac{\xi}{2eB}\SB{\frac{}{}\sper-2\kper^2-2\pper^2}}} 
\nn \\ && \hspace{0cm} \times
\int_{0}^{1}\!\!dy\TB{\Theta(Z_+)\Theta(1-y-Z_+)+\Theta(Z_-)\Theta(1-y-Z_-)}\mathcal{T}_D^\Strong(\spll,\kper^2,\pper^2) \label{eq.impiD.strong}
\end{eqnarray}
where, $Z^\pm = \Ztil_\pm(\lambda=0)$ which can be obtained from Eq.~\eqref{eq.Z}; 
$\mathcal{T}_B^\Strong$ and $\mathcal{T}_D^\Strong$ is given in Eqs.~\eqref{eq.TB.Strong} and \eqref{eq.TD.Strong}.

Thus, we have obtained the imaginary parts of the forward scattering amplitudes under external magnetic field which 
can be substituted in Eq.~\eqref{eq.optical.eB} to evaluate the cross-section using the optical theorem.

\section{RESULTS \& DISCUSSIONS}\label{sec.results}
Having calculated the cross-section for elastic $b^+(k)b^-(p)\rightarrow b^+(k')b^-(p')$ scattering in the previous sections, we now proceed to show some numerical results for our analysis. We have chosen the values of the masses of $B^0$ and $b^\pm$ as $M=400$ MeV and $m=300$ MeV respectively for the numerical estimation of cross-section. We begin this section by showing the cross-section at \textit{zero external magnetic field}. At $B=0$, the vacuum cross-section is obtained either directly from tree graphs using Eq.~\eqref{eq.croos.tree} or from the loop graphs through the optical theorem using Eq.~\eqref{eq.optical}. 
\begin{figure}[h]
\begin{center}
\includegraphics[angle=-90, scale=0.35]{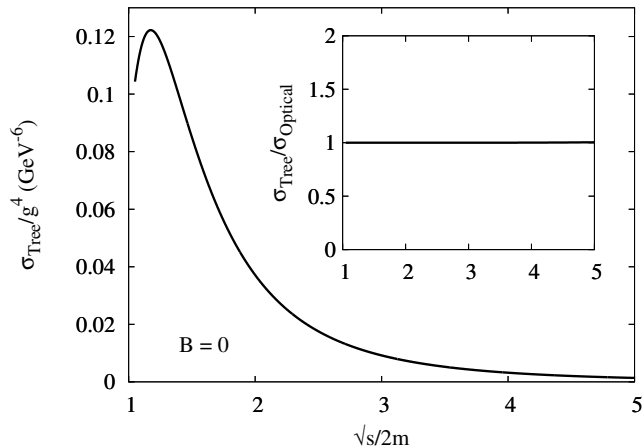}
\end{center}
\caption{The variation of vacuum cross-section calculated directly from the tree graphs scaled with fourth power of 
	the inverse coupling as a function of scaled center of mass energy ($\sqrt{s}/2m$). 
	The inset plot shows the ratio of the cross-sections obtained from tree graphs 
to that from one-loop graphs using optical theorem as a function of scaled center of mass energy.}
\label{fig.CrossVac}
\end{figure}
We denote the vacuum cross-section obtained from the tree graphs by $\sigma_\text{Tree}(s)$ whereas the same obtained from the loop graphs through the optical theorem by $\sigma_\text{Optical}(s)$. Since the cross-section is proportional to the fourth power of the coupling constant $g$, we have plotted the ratio $\sigma_\text{Tree}(s)/g^4$ as a function of scaled center of mass energy $\sqrt{s}/2m$ in Fig.~\ref{fig.CrossVac}. With the increase in $\sqrt{s}$, the cross-section decreases, which is due to the suppression of the $s$-channel contributions to the invariant amplitude at higher $\sqrt{s}$.

As already mentioned at the end of Sec.~\ref{sec.loop}, the cross-section obtained from the loop graphs through optical theorem $\sigma_\text{Optical}(s)$ should agree with the $\sigma_\text{Tree}(s)$. To show that, we have plotted the ratio $\sigma_\text{Tree}/\sigma_\text{Optical}$ as a function of $\sqrt{s}/2m$ in the inset plot of Fig.~\ref{fig.CrossVac}. As it turns out from the figure, that the ratio is always unity implying 
\begin{eqnarray}
\sigma_\text{Optical}(s) = \sigma_\text{Tree}(s) = \sigma_0(s)
\end{eqnarray}
where `0' in the subscript denotes the vacuum cross-section i.e at $B=0$.

We now consider the case of \textit{non-zero external magnetic field}. In this case we have calculated the cross-section from the optical theorem for the two cases separately: (a) at weak field approximation and (b) at strong field approximation. We denote them respectively by $\sigma_\Weak$ and $\sigma_\Strong$. Unlike the $B=0$ case, where the cross-section depends only on the total center of mass energy $\sqrt{s}$, for $B\ne0$, the cross-section additionally depends on three more Lorentz scalars $\sper=(\kper+\pper)^2$, $\kper^2$ and $\pper^2$. These three are related to the quantities $\kperv$, $\pperv$ and $\theta$ with $\theta$ being the angle between $\vec{k}_\perp$ and $\vec{p}_\perp$. We have expressed the cross-section as a function of $\sqrt{s}$, $\kperv$, $\pperv$ and $\theta$ for presenting the numerical results.
\begin{figure}[h]
	\begin{center}
		\includegraphics[angle=-90, scale=0.35]{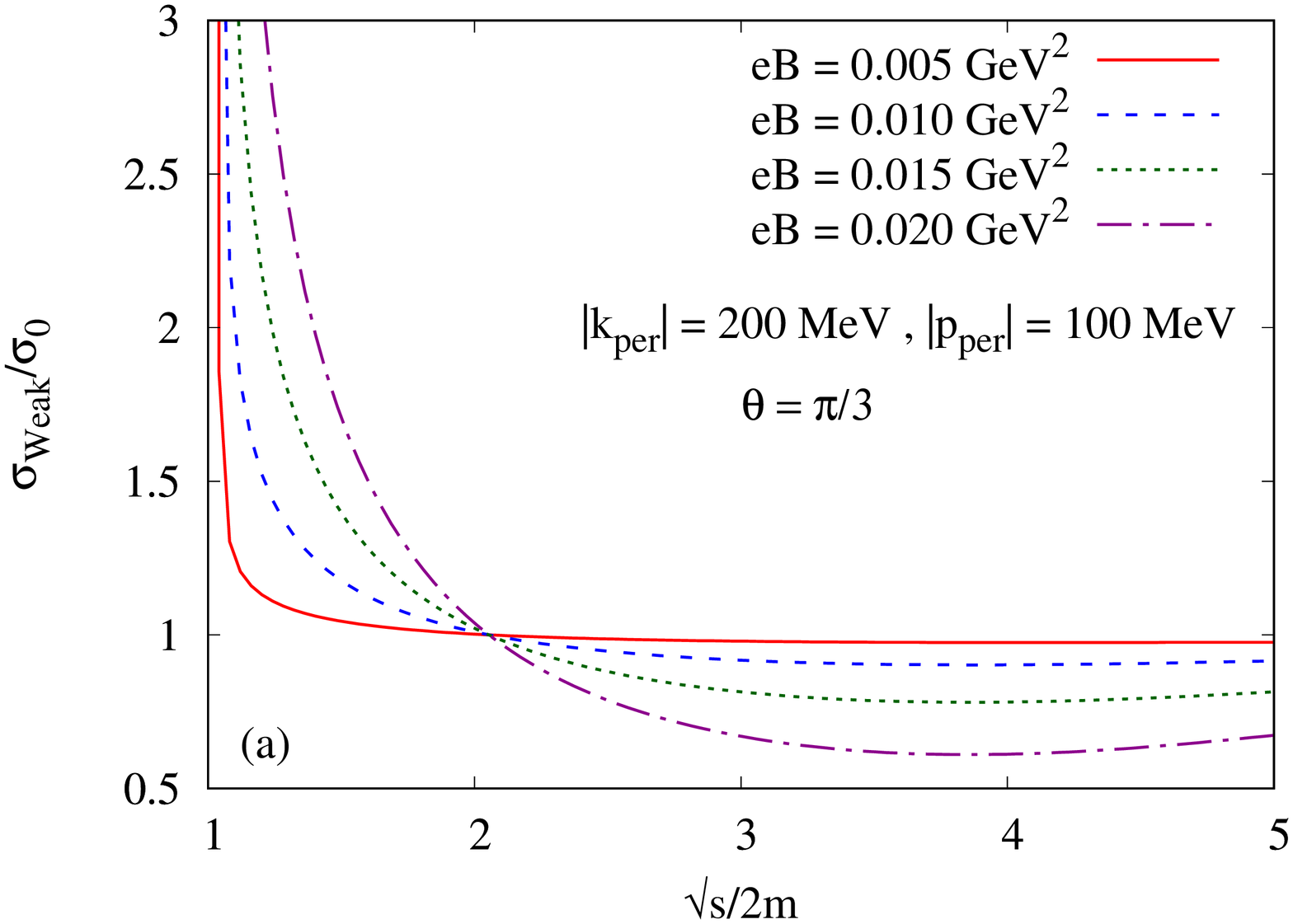}
		\includegraphics[angle=-90, scale=0.35]{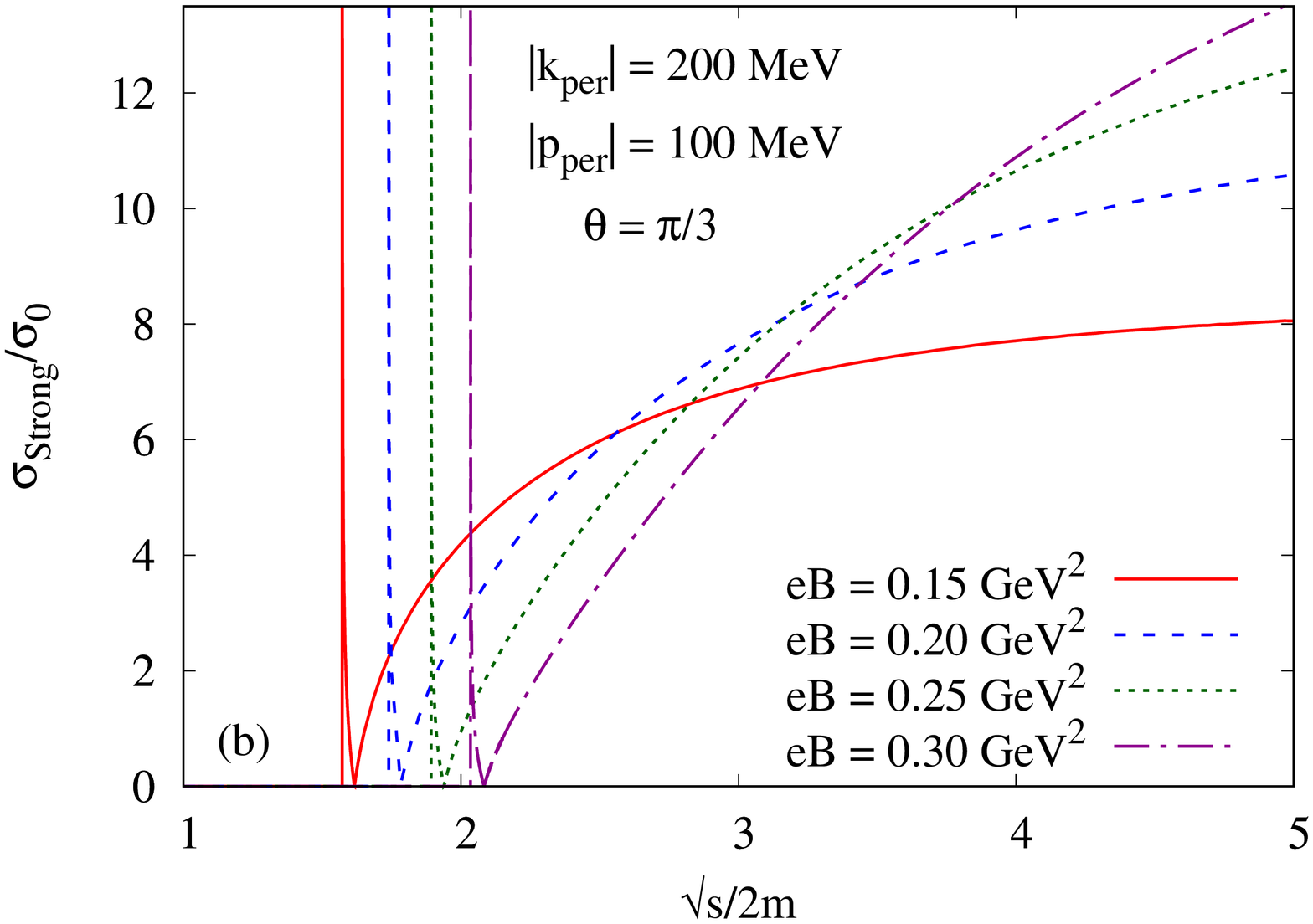}		
	\end{center}
\caption{The ratio of the cross-section under (a) weak and (b) strong external magnetic field to the vacuum cross-section as a function of scaled center of mass energy ($\sqrt{s}/2m$) at $\kperv=200$ MeV, $\pperv=100$ MeV, $\theta=\pi/3$. Subfigure (a) shows the results at four different lower values of magnetic field ($eB=0.005, 0.010, 0.015$ and $0.020$ GeV$^2$ respectively) valid for the weak field approximation whereas (b) shows the same at four different higher values of magnetic field ($eB=0.15, 0.20, 0.25$ and $0.30$ GeV$^2$ respectively) valid for the strong field approximation.}
	\label{fig.Cross.1}
\end{figure}

In Fig.~\ref{fig.Cross.1}(a), we have shown the ratio $\sigma_\Weak/\sigma_0$ as a function of $\sqrt{s}/2m$ at $\kperv=200$ MeV, $\pperv=100$ MeV, $\theta=\pi/3$ and at four different values of external magnetic field ($eB=0.005, 0.010, 0.015$ and $0.020$ GeV$^2$ respectively). The strengths of the magnetic fields are chosen in such a way that the validity of the weak field approximation $eB\ll m^2$ holds. As can be seen from the graph that, around $\sqrt{s}=2m$ (which is the threshold center of mass energy for the scattering process) we have huge enhancement of the cross-section with respect to the vacuum. This is due to the presence of `threshold singularity' as can be understood from Eqs.~\eqref{eq.impiA.weak}, \eqref{eq.impiBC.weak} and \eqref{eq.impiD.weak} in which the weak field corrections to the imaginary part of the amplitudes goes like $\simeq \frac{1}{(s-4m^2)^\eta}$ with $\eta>0$. The threshold being $\sqrt{s}=2m$, the cross-section sufferers from `threshold singularity'. The enhancement of the cross-section due to external magnetic field is more at higher values of magnetic field. With the increase in $\sqrt{s}$, the ratio $\sigma_\Weak/\sigma_0$ decreases and eventually it becomes less than unity at $\sqrt{s}\gtrsim 4m$ owing to a saturating behaviour at higher $\sqrt{s}$. 
Although the weak field corrections induced by the magnetic field effects go as $(eB)^2$, yet the correction to the cross-section due to the magnetic field at high $\sqrt{s}$ comes out to be negative (as $\sigma_\Weak < \sigma_0$). This is due to the fact that, in Eqs.~\eqref{eq.impiA.weak}-\eqref{eq.impiD.weak}, the coefficients of $(eB)^2$ can have either signs depending on the values of $s$. This makes the cross-section to be enhanced or decreased with respect to the vacuum cross-section depending on $\sqrt{s}$. Moreover, as the $s$-channel contribution to the cross-section is suppressed at high center of mass energy, the dominant contribution to the cross-section at high $\sqrt{s}$ comes from the $t$-channel (i.e. the term $\IM\Pi^\text{Weak}_D$ in Eq.~\eqref{eq.impiD.weak}). Since the quantity $\mcT^\text{Weak}_D$ is positive in the region $\sqrt{s}\gg m$, the weak field cross-section $\sigma_\text{Weak}$ becomes less than the vacuum cross-section $\sigma_0$ at high center of mass energy.

The corresponding behaviour of the cross-section in the strong field approximation is shown in Fig.~\ref{fig.Cross.1}(b) where we have plotted the ratio $\sigma_\Strong/\sigma_0$ as a function of $\sqrt{s}/2m$ at $\kperv=200$ MeV, $\pperv=100$ MeV, $\theta=\pi/3$ and at four different values of external magnetic field ($eB=0.15, 0.20, 0.25$ and $0.030$ GeV$^2$ respectively). The strengths of the magnetic fields are chosen to satisfy $eB\gg m^2$ which is the necessary requirement for the validity of the strong field approximation. In this case, one can see that the threshold of the cross-section depends on the external magnetic field and it moves towards higher values of $\sqrt{s}$ with the increase in magnetic field. There is also a huge enhancement of the cross-section due to the external magnetic field and the enhancement is more a higher $\sqrt{s}$. Moreover, at the threshold, there are a large spikes which is again due to the `threshold singularity' in the lowest Landau level. The behaviour of the plot can be understood from Eqs.~\eqref{eq.impiA.strong}, \eqref{eq.impiBC.strong} and \eqref{eq.impiD.strong} in which the imaginary parts of the amplitudes goes like $\simeq \frac{1}{(\spll-4m^2-4eB)^\eta}$ with $\eta>0$. The threshold being $\spll\ge4(m^2+eB)$ thus moves towards higher $\sqrt{s}$ with an increase in $eB$. Similar kind of spike like behaviour in the scattering cross-section is also present in an earlier work~\cite{Bhattacharya:2002qf}. 
\begin{figure}[h]
	\begin{center}
		\includegraphics[angle=-90, scale=0.35]{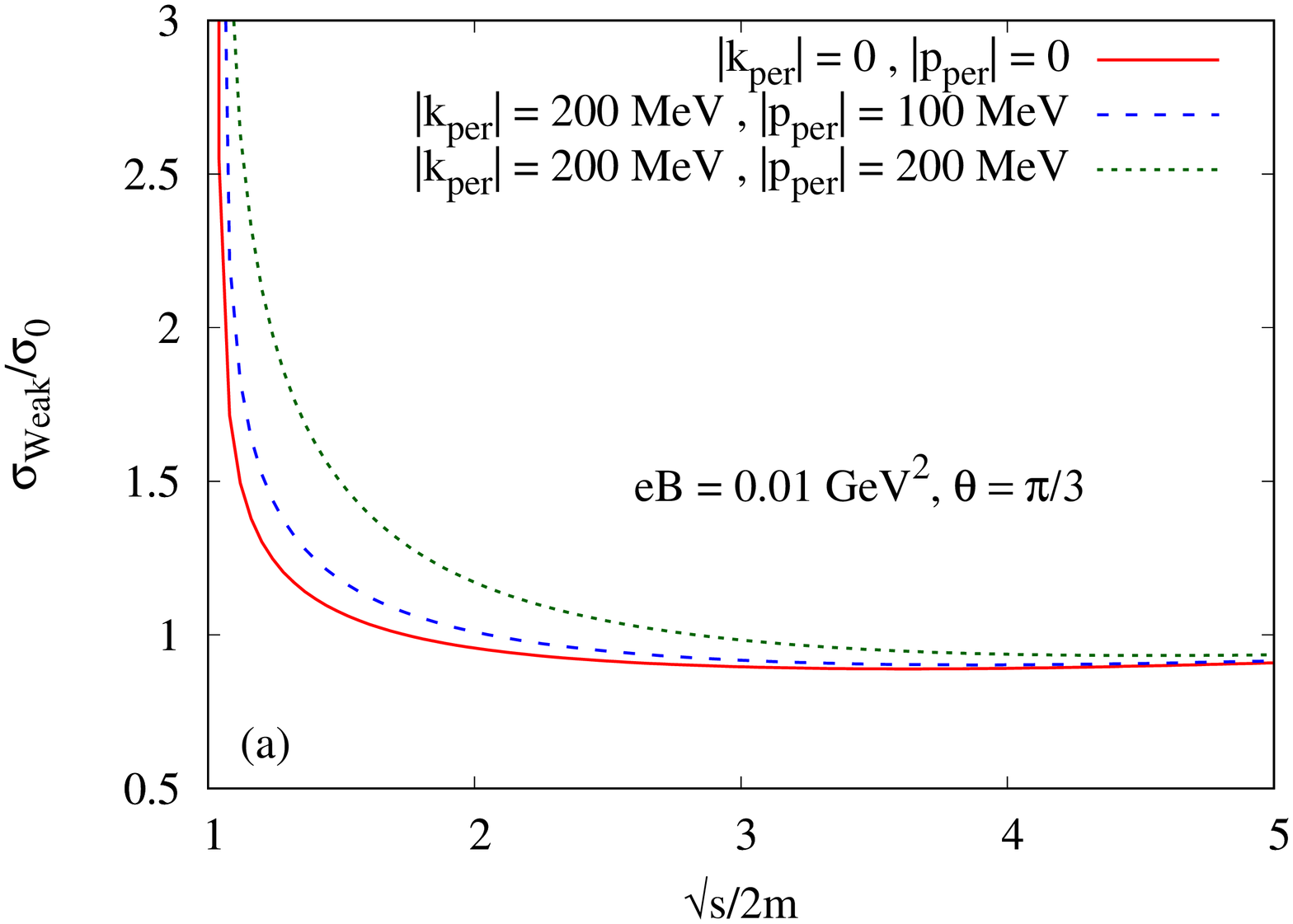}
		\includegraphics[angle=-90, scale=0.35]{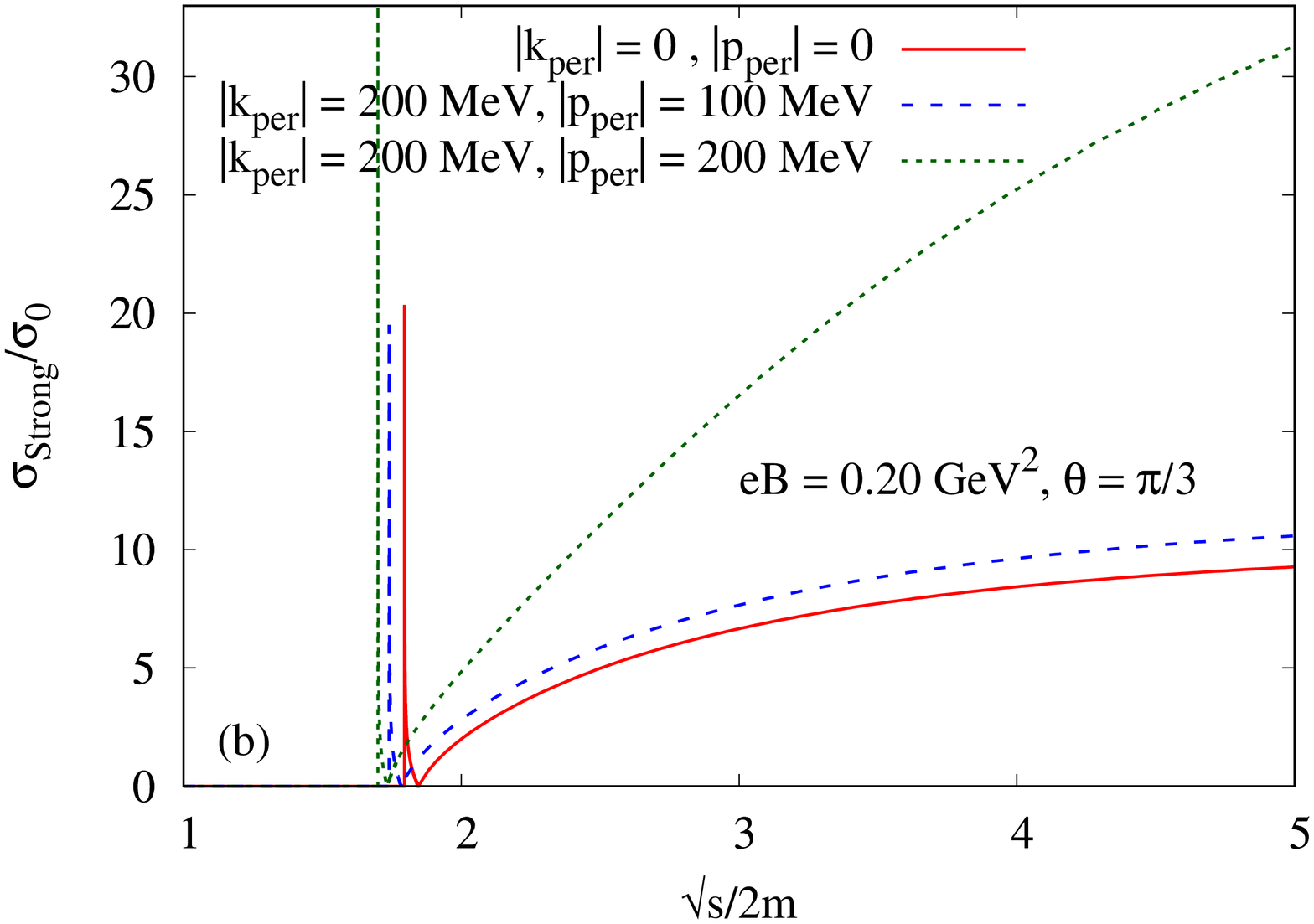}		
	\end{center}
\caption{The ratio of the cross-section under (a) weak and (b) strong external magnetic field to the vacuum cross-section as a function of 
	scaled center of mass energy ($\sqrt{s}/2m$) at $\theta=\pi/3$ and at different combinations of $\kperv$ and $\pperv$. 
	Subfigure (a) shows the results at a lower value of magnetic field ($eB=0.01$ GeV$^2$) valid for the weak field 
	approximation whereas (b) shows the same at a higher value of magnetic field ($eB=0.20$ GeV$^2$) valid for the strong field approximation.}
\label{fig.Cross.2}
\end{figure}
\begin{figure}[h]
	\begin{center}
		\includegraphics[angle=-90, scale=0.35]{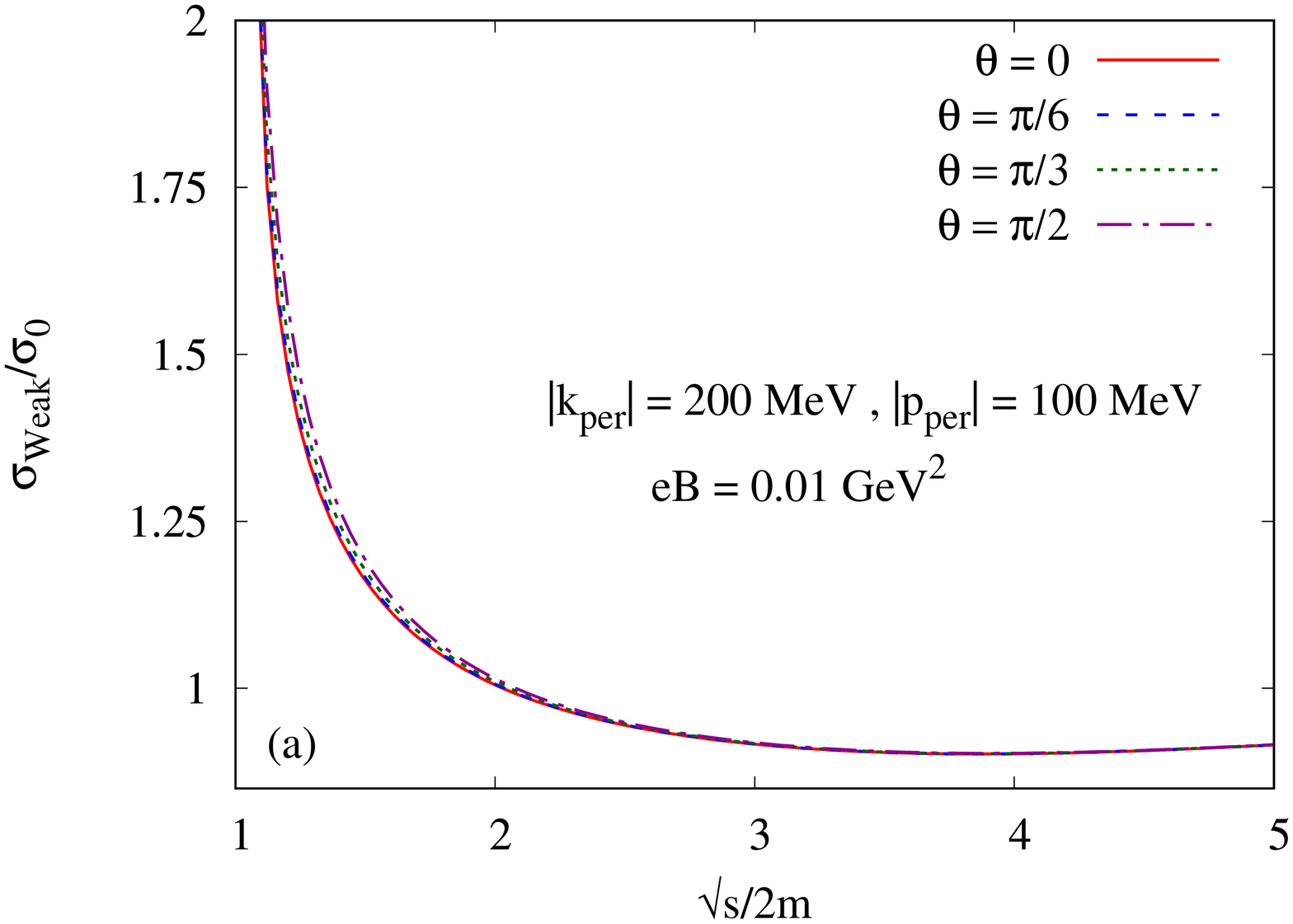}
		\includegraphics[angle=-90, scale=0.35]{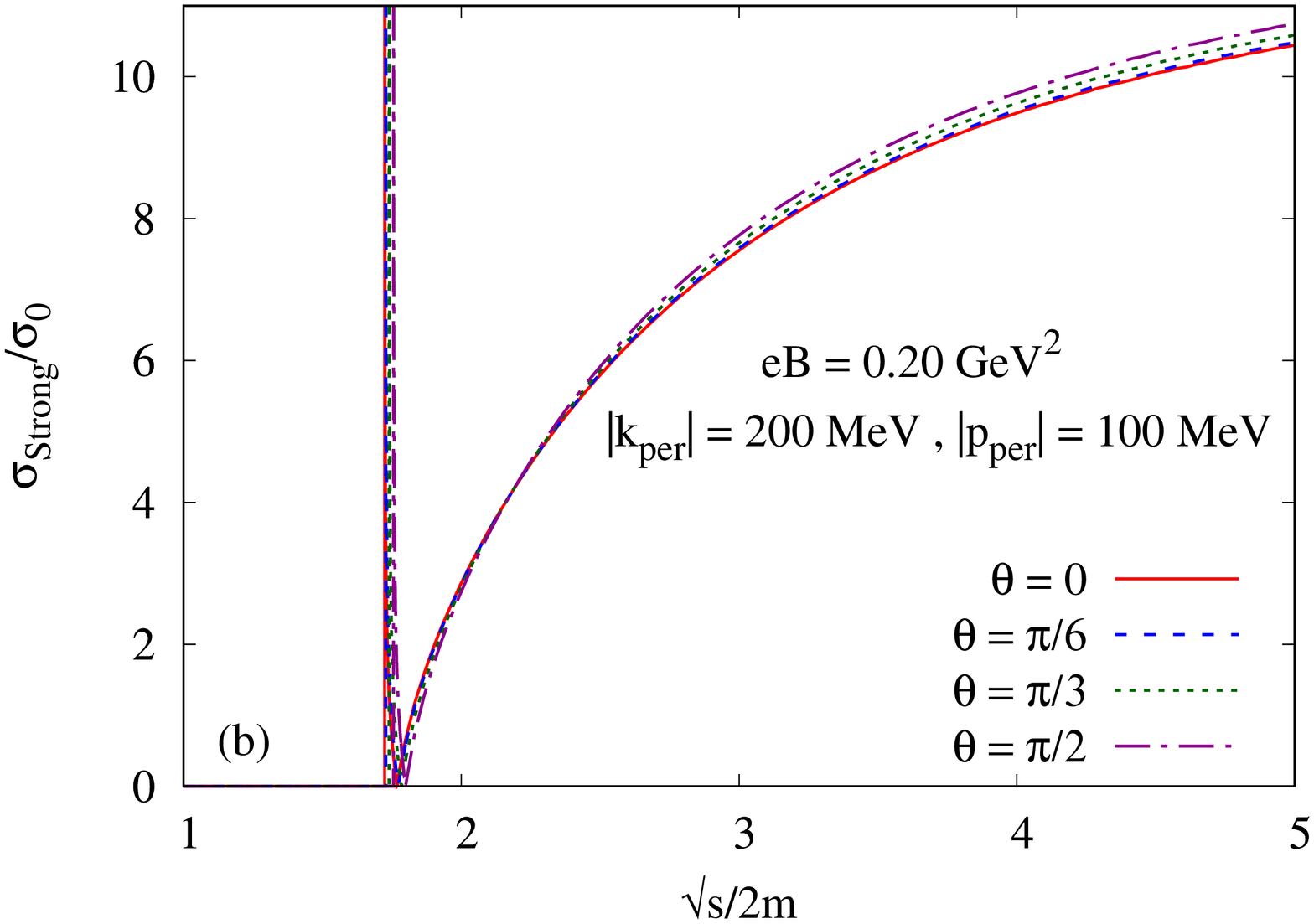}		
	\end{center}
\caption{The ratio of the cross-section under (a) weak and (b) strong external magnetic field to the vacuum cross-section as a function of 
	scaled center of mass energy ($\sqrt{s}/2m$) at $\kperv=200$ MeV, $\pperv=100$ MeV and four different values of 
	$\theta (0,\pi/6,\pi/3$ and $\pi/2$ respectively).	Subfigure (a) shows the results at a lower value of magnetic 
	field ($eB=0.01$ GeV$^2$) valid for the weak field approximation whereas (b) shows the same at a higher value of magnetic 
	field ($eB=0.20$ GeV$^2$) valid for the strong field approximation.}
\label{fig.Cross.3}
\end{figure}

Having studied the variation of the cross-section with the external magnetic field, we now proceed to show its variation with the other parameters namely $\kperv$, $\pperv$ and $\theta$. Fig.~\ref{fig.Cross.2}(a), we have plotted the ratio $\sigma_\Weak/\sigma_0$ as a function of $\sqrt{s}/2m$ at $eB=0.01$ GeV$^2$, $\theta=\pi/3$ and at three different combinations for $\kperv$ and $\pperv$. In all the cases the cross-section is enhanced with respect to the vacuum cross-section and the enhancement is more at higher values of transverse momenta.

The corresponding graph in the strong field approximation is shown in Fig.~\ref{fig.Cross.2}(b) where the ratio $\sigma_\Strong/\sigma_0$ is plotted as a function of $\sqrt{s}/2m$ at $eB=0.20$ GeV$^2$, $\theta=\pi/3$ and at three different combinations for $\kperv$ and $\pperv$. In this case, the threshold of the cross-section defined in terms of 
\begin{eqnarray}
\spll\ge4(m^2+4eB)~~\text{or}~~s\ge4(m^2+eB)-(\kperv^2+\pperv^2+2\kperv\pperv\cos\theta)~,
\label{eq.Strong.Threshold}
\end{eqnarray}
thus depends on the transverse momenta and it moves towards lower values of $\sqrt{s}$ with the increase in $\kperv$ and $\pperv$. Analogous to the weak field case, the enhancement of cross-section is more at a higher transverse momenta.

Next, we study the dependence of the cross-section under external magnetic field on the parameter $\theta$. Fig.~\ref{fig.Cross.3}(a) shows the variation of the ratio $\sigma_\Weak/\sigma_0$ as a function of $\sqrt{s}/2m$ at $eB=0.01$ GeV$^2$, $\kperv=200$ MeV, $\pperv=100$ MeV and at four different values of $\theta~(0,\pi/6,\pi/3$ and $\pi/2$). It can be seen from the graph that the dependence of $\sigma_\Weak$ on $\theta$ is weak and small variation is observed at low $\sqrt{s}$ region.

Analogous plot for the case of strong field approximation is shown in Fig.~\ref{fig.Cross.3}(b) where $\sigma_\Strong/\sigma_0$ is shown as a function of $\sqrt{s}/2m$ at $eB=0.20$ GeV$^2$, $\kperv=200$ MeV, $\pperv=100$ MeV and at four different values of $\theta~(0,\pi/6,\pi/3$ and $\pi/2$). In this case, the threshold of the cross-section also depends on $\theta$ as can be obtained from Eq.~\eqref{eq.Strong.Threshold}. The threshold energy, being minimum at $\theta=0$, increases with the increase in $\theta$ and will reach maximum at $\theta=\pi$. The similar behaviour can be noticed in the figure. Unlike the weak field case, the dependence of the cross-section on theta is more at higher values of $\sqrt{s}$.
\begin{figure}[h]
	\begin{center}
		\includegraphics[angle=-90, scale=0.35]{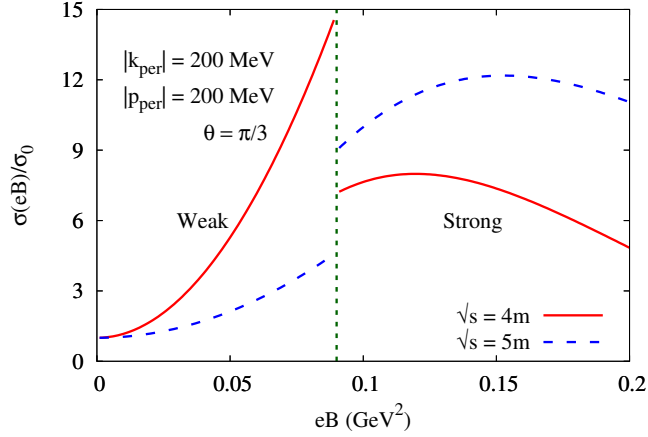}
	\end{center}
\caption{The ratio of the cross-section under external magnetic field to the vacuum cross-section as a function of external magnetic field 
	at $\kperv=200$ MeV, $\pperv=100$ MeV, $\theta=\pi/3$ and at two different values of the center of mass energies 
	($\sqrt{s}=4m$ and $5m$ respectively). The vertical green line corresponds to $eB=m^2$ which separates the region of validity of the 
	weak and strong field approximation. The cross-section in regions $eB<m^2$ and $eB>m^2$ are obtained using the weak and strong field approximations respectively.}
\label{fig.Cross.4}
\end{figure}

Since, we have calculated the cross-section employing the weak and strong field approximations, we thus can not rely on them for the intermediate values of the external magnetic field. The weak field approximation is valid for $eB\ll m^2$ whereas the validity of strong field approximation is $eB\gg m^2$. Thus $eB=m^2$ is the boundary above (below) which one can not consider the weak (strong) field approximation. In Fig.~\ref{fig.Cross.4}, we have plotted the variation of ratio $\sigma(eB)/\sigma_0$ as a function of $eB$ at $\kperv=\pperv=200$ MeV, $\theta=\pi/3$ and at two different values of center of mass energy ($\sqrt{s}=4m$ and $5m$ respectively). The whole $eB$ axis is divided into two regions: $eB<m^2$ and $eB>m^2$ by the green (dash-dot) line. The cross-section in the former region is obtained employing the weak field approximation whereas in the later, we use the strong field approximation. It can be seen from the graph that, there is a discontinuity at $eB=m^2$ which is obvious due to the fact that, around $eB\simeq m^2$, both the approximations break down. It is worth noting that, incorporation of both the higher order corrections ($\mathcal{O}(eB)^4$ etc.) to the $\sigma_\Weak$ and contributions from few more higher Landau levels to the $\sigma_\Strong$ will fine tune our estimate of cross-section. However, a more comprehensive picture around the region $eB \simeq m^2$ could be obtained only if we extend our calculation considering the full Schwinger propagator including all the Landau levels. These issues are beyond the scope of the present work and will be taken up as the immediate future work.

\section{SUMMARY \& CONCLUSIONS} \label{sec.summary}
In summary, we have calculated the cross-section for the elastic scattering between two charged scalars mediated via a neutral scalar under an external magnetic field within the optical theorem. The forward scattering amplitudes are calculated using standard field theoretic techniques in which the effect of external magnetic field entered through the modification of the charged scalar propagators by the Schwinger proper-time one. Calculations on the cross-section employing both the weak as well as strong magnetic field approximations have been performed.

It is observed that, the cross-section obtained from the optical theorem at zero external magnetic field reproduces the same obtained using the tree-level diagrams. For the weak field approximation, the cross-section is enhanced with respect to the vacuum cross-section at lower center of mass energies whereas for the strong field approximation the cross-section is enhanced with respect to the vacuum cross-section at all the values of center of mass energies. Unlike the vacuum cross-section, which depends only on the center of mass energy (the Lorentz scalar $s$ of Mandelstam variables), the cross-section under the eternal magnetic field depends on four Lorentz scalars which are related to the magnitude and orientations of the transverse momenta of the incoming particle with respect to the direction of the external magnetic field. The cross-section under external magnetic field also has a stronger dependence on these parameters in addition to the total center of mass energy. Finally, the cross-section has also a nontrivial magnetic field dependence. As shown in Fig.~\ref{fig.Cross.4}, the enhancement of the cross-section due to the external magnetic field has a qualitative agreement with Ref.~\cite{Bhattacharya:2002qf} calculated for the inverse beta decay process.

The possible improvement over this work could be the incorporation of $\mathcal{O}(eB)^4$ contributions to $\sigma_\Weak$ as well as the incorporation of few more Landau levels to the $\sigma_\Strong$. This will fine tune our estimate of the cross-section. The incorporation of finite temperature in addition to the external magnetic field is also an interesting problem to carry out. In particular, this will be relevant for the study of hot and dense matter created in HIC experiments. As we have already discussed, the various transport coefficients (viscosities and conductivities) and the nuclear modification factor (obtained from drag and diffusion coefficients of heavy quarks) of the magnetized hot and dense ``strongly" interacting matter created in HIC experiments will primarily be affected by the modification in the scattering cross-sections due to the external magnetic field. These aspects will be the matter of future investigations.

\section*{Acknowledgments}
S.G. acknowledges Arghya Mukherjee for useful discussions. 
S.G. also acknowledges the Indian Institute of Technology Gandhinagar, Saha Institute of Nuclear Physics, Kolkata 
and the Department of Atomic Energy (DAE), Government of India for financial support. 
V.C. would like to acknowledge SERB, Government of India for Early Career Research Award (ECRA/2016/000683) 
and INSA-Department of Science and Technology, Government of India for INSPIRE Faculty award (IFA-13/PH-55). 
The people of India are sincerely acknowledged for their invaluable support for the research in fundamental sciences.

\appendix

\section{USEFUL IDENTITIES} \label{appendix.identities}
We have the following list of $d$-dimensional integrals in Minkowski space~\cite{Peskin:1995ev}:
\begin{eqnarray}
\int\!\!\frac{d^dk}{\FB{2\pi}^d}\frac{1}{\FB{k^2-\Delta}^n} &=& 
\frac{i\FB{-1}^n}{\FB{4\pi}^{d/2}}\frac{\Gamma\FB{n-d/2}}{\Gamma\FB{n}}\FB{\frac{1}{\Delta}}^{n-d/2}, \\
\int\!\!\frac{d^dk}{\FB{2\pi}^d}\frac{k^2}{\FB{k^2-\Delta}^n} &=& 
\frac{i\FB{-1}^{n-1}}{\FB{4\pi}^{d/2}}\FB{\frac{d}{2}}\frac{\Gamma\FB{n-1-d/2}}{\Gamma\FB{n}}\FB{\frac{1}{\Delta}}^{n-1-d/2}, \\
\int\!\!\frac{d^dk}{\FB{2\pi}^d}\frac{k^\mu k^\nu}{\FB{k^2-\Delta}^n} &=& 
\frac{i\FB{-1}^{n-1}}{\FB{4\pi}^{d/2}}\FB{\frac{g^\munu}{2}}\frac{\Gamma\FB{n-1-d/2}}{\Gamma\FB{n}}\FB{\frac{1}{\Delta}}^{n-1-d/2}.
\end{eqnarray}

The following integral representation of the modified Bessel function of the first kind $I_0(z)$ is required for the 
calculation of the amplitudes in the strong field approximation:
\begin{eqnarray}
I_0(z) = \int_{0}^{2\pi}\!\frac{d\phi}{2\pi}e^{-z\cos\phi}.
\end{eqnarray}


\section{IMAGINARY PARTS OF THE FORWARD SCATTERING AMPLITUDES IN THE VACUUM} \label{appendix.impiABCD.vac}
In this appendix, we will sketch the calculations of the imaginary parts of different forward scattering amplitudes in the vacuum 
\textit{i.e. in the absence of external magnetic field} as given in Eqs.~\eqref{eq.A.vac}-\eqref{eq.D.vac}.

Considering the imaginary part of Eq.~\eqref{eq.A.vac}, we get
\begin{eqnarray}
	\IM\Pi_A^\vac(k,p) = g^2\TB{\frac{\IM\Pi_\vac(k+p)}{(s-M^2)^2} - 2\pi\delta(s-M^2)\frac{\RE\Pi_\vac(k+p)}{(s-M^2)}}~. \label{eq.ImA.1}
\end{eqnarray}
We now expand $\Pi_\vac(k+p)$ as
\begin{eqnarray}
	\Pi_\vac(k+p) = \Pi_\vac(s=M^2) + (s-M^2)\frac{d\Pi_\vac}{ds}\Bigg|_{s=M^2} + \mathcal{O}(s-M^2)^2~. \label{eq.PiExpansion}
\end{eqnarray}
Using the fact that, $\RE\Pi_\vac$ renormalizes the bare $B^0$ mass to its physical mass, the renormalization conditions require~\cite{Peskin:1995ev}
\begin{eqnarray}
	\Pi_\vac(s=M^2) = 0 ~~~ \text{and} ~~~~ \frac{d\Pi_\vac}{ds}\Big|_{s=M^2} =0~. \label{eq.Renormalization}
\end{eqnarray}
Substitution of Eqs.~\eqref{eq.PiExpansion} and Eq.~\eqref{eq.Renormalization} into Eq.~\eqref{eq.ImA.1} would imply that, the second term 
within the square bracket in Eq.~\eqref{eq.ImA.1} will not contribute and we are left with
\begin{eqnarray}
	\IM\Pi_A^\vac(k,p) = g^2\TB{\frac{\IM\Pi_\vac(k+p)}{(s-M^2)^2}}~. \label{eq.ImA.2}
\end{eqnarray}
The calculation of $\IM\Pi_\vac$ is provided in Appendix~\ref{appendix.Pi.Vac} and we get from Eq.~\eqref{eq.ImPi.Vac}
\begin{eqnarray}
	\IM\Pi_\vac(k+q) = -\frac{g^2}{16\pi s}\sqrt{s-4m^2}\Theta(s-4m^2) 
\end{eqnarray}
so that, Eq.~\eqref{eq.ImA.2} finally becomes
\begin{eqnarray}
	\IM\Pi_A^\vac(s) = -\frac{g^4}{16\pi s(s-M^2)^2}\sqrt{s-4m^2}\Theta(s-4m^2)~. \label{eq.ImA.Final}
\end{eqnarray}
It can be seen that, the $\IM\Pi_A^\vac(k,p)$ depends only on the Lorentz scalar $s=(k+p)^2$.


Next we proceed to calculate $\IM\Pi_B^\vac(k,p)$ from Eq.~\eqref{eq.B.vac} which is given by,
\begin{eqnarray}
	\IM\Pi_B^\vac(k,p)=g\TB{\frac{-\IM\mcV_\vac(k,p)}{(s-M^2)}+\pi\delta(s-M^2)\RE\mcV_\vac(k,p)}~.
\end{eqnarray}
Using the fact that, $\mcV_\vac(k,p)$ renormalize the bare coupling to the physical coupling, the renormalization 
condition requires~\cite{Peskin:1995ev} 
$\mcV_\vac(k,p)\Big|_{(k+p)^2=M^2}=0$, which in turn makes the second term within square bracket in above equation zero. Thus 
$\IM\Pi_B^\vac(k,p)$ becomes
\begin{eqnarray}
	\IM\Pi_B^\vac(k,p)=g\TB{\frac{-\IM\mcV_\vac(k,p)}{(s-M^2)}}~. \label{eq.ImB.Vac}
\end{eqnarray}
The calculation of $\IM\mcV_\vac$ is provided in Appendix~\ref{appendix.Vertex.Vac} and we get from Eq.~\eqref{eq.Im.Vertex.Vac.1}
\begin{eqnarray}
	\IM\mcV_\vac(k,p) &=& \frac{-g^3}{16\pi}\int_{0}^{1}\!\!dy\frac{\Theta(z_+)\Theta(1-y-z_+)+\Theta(z_-)\Theta(1-y-z_-)}
	{\sqrt{(M^2+sy)^2-4m^2(M^2+sy^2)}}~, \label{eq.Im.Vertex.Vac.2}
\end{eqnarray}
with $z_\pm$ is obtained from Eq.~\eqref{eq.zpm} as
\begin{eqnarray}
	z_\pm(s,y) = \frac{1}{2m^2}\TB{M^2-2m^2y+sy\pm\sqrt{(M^2+sy)^2-4m^2(M^2+sy^2)}}~. \label{eq.zpm.1}
\end{eqnarray}
Substituting Eq.~\eqref{eq.Im.Vertex.Vac.2} into Eq.~\eqref{eq.ImB.Vac}, we finally get
\begin{eqnarray}
	\IM\Pi_B^\vac(s) = \frac{g^4}{16\pi(s-M^2)}\int_{0}^{1}\!\!dy\frac{\Theta(z_+)\Theta(1-y-z_+)+\Theta(z_-)\Theta(1-y-z_-)}
	{\sqrt{(M^2+sy)^2-4m^2(M^2+sy^2)}}~. \label{eq.ImB.Final}
\end{eqnarray}
As it turns out that the $\IM\Pi_B^\vac(k,p)$ is a function of the Lorentz scalar $s=(k+p)^2$ only, it immediately follows from 
Eq.~\eqref{eq.C.vac} that
\begin{eqnarray}
	\IM\Pi_C^\vac(s) = \IM\Pi_B^\vac(s)~. \label{eq.ImC.Final}
\end{eqnarray}

Finally, we have from Eq.~\eqref{eq.D.vac},
\begin{eqnarray}
	\Pi_D^\vac(k,p) &=& ig^4\!\int\!\!\frac{d^4\ktilde}{(2\pi)^4}\Delta^2_F(\ktilde,M)\Delta_F(k-\ktilde,m)\Delta_F(p+\ktilde,m) \nn \\
	&=& ig^4\!\int\!\!\frac{d^4\ktilde}{(2\pi)^4}\frac{1}{{(\ktilde^2-M^2+i\epsilon)^2\{(k-\ktilde)^2-m^2+i\epsilon\}\{(p+\ktilde)^2-m^2+i\epsilon\}}}~.
\end{eqnarray}
Using Feynman parametrization, we combine the denominator of the above equation and get,
\begin{eqnarray}
	\Pi_D^\vac(k,p) &=& 6ig^4\!\int\!\!\frac{d^4\ktilde}{(2\pi)^4}\int_{0}^{1}\!\!\int_{0}^{1}\!\!\int_{0}^{1}\!\!dxdydz
	\delta(1-x-y-z)\frac{(1-y-z)}{[(\ktilde-yk+zp)^2-\Delta_\mcV]^4} 
\end{eqnarray}
where $\Delta_\mcV$ is defined in Eq.~\eqref{eq.Delta.V.Vac.2} as
\begin{eqnarray}
	\Delta_\mcV = (y+z)^2m^2+(1-y-z)M^2-yzs-i\varepsilon~.
\end{eqnarray}
Next we shift $\ktilde\rightarrow(\ktilde+yk-zp)$ in the above equation and perform the momentum integration using the identities provided in Appendix~\ref{appendix.identities} to obtain
\begin{eqnarray}
	\Pi_D^\vac(k,p) &=& \frac{-g^4}{16\pi^2}\!\int_{0}^{1}\!\!\int_{0}^{1}\!\!\int_{0}^{1}\!\!dxdydz
	\delta(1-x-y-z)\frac{(1-y-z)}{\Delta_\mcV^2}~.
\end{eqnarray}
The $dx$ integral in the above equation is performed using the Dirac delta function present in the integrand and we get,
\begin{eqnarray}
	\Pi_D^\vac(k,p) &=& \frac{-g^4}{16\pi^2}\!\int_{0}^{1}\!\!dy\int_{0}^{1-y}\!\!dz\frac{(1-y-z)}{\Delta_\mcV^2}~.
\end{eqnarray}
The presence of small negative imaginary part in $\Delta_\mcV$ will give rise to the imaginary part of the amplitude in 
certain kinematic domains. In order to calculate the imaginary part of $\Pi_D^\vac$, we use the following trick :
\begin{eqnarray}
	\IM\TB{\frac{1}{(x-i\epsilon)^n}} = 
	\IM\TB{\frac{(-1)^{n-1}}{(n-1)!}\frac{\partial^{n-1}}{\partial\lambda^{n-1}}\FB{\frac{1}{x+\lambda-i\epsilon}}}\Bigg|_{\lambda=0} 
	= \pi \frac{(-1)^{n-1}}{(n-1)!}\frac{\partial^{n-1}}{\partial\lambda^{n-1}}\delta(x+\lambda)\Bigg|_{\lambda=0} \label{eq.Trick}
\end{eqnarray}
and write
\begin{eqnarray}
	\IM\Pi_D^\vac(k,p) &=& \frac{g^4}{16\pi}\frac{\partial}{\partial\lambda}\int_{0}^{1}\!\!dy\int_{0}^{1-y}\!\!dz(1-y-z)
	\delta\TB{(y+z)^2m^2+(1-y-z)M^2-yzs+\lambda}\Bigg|_{\lambda=0}~. \label{eq.D.vac.1}
\end{eqnarray}
To simplify the above expression, we transform the Dirac delta function as 
\begin{eqnarray}
	\delta\TB{(y+z)^2m^2+(1-y-z)M^2-yzs+\lambda} = \frac{\delta(z-\ztil_+)+\delta(z-\ztil_-)}{\sqrt{(M^2+sy)^2-4m^2(M^2+sy^2+\lambda)}}
	\label{eq.delta.2}
\end{eqnarray}
where,
\begin{eqnarray}
	\ztil_\pm(s,y,\lambda) = \frac{1}{2m^2}\TB{M^2-2m^2y+sy\pm\sqrt{(M^2+sy)^2-4m^2(M^2+sy^2+\lambda)}}~. \label{eq.ztilpm.1}
\end{eqnarray}
Substituting Eq.~\eqref{eq.delta.2} into Eq.~\eqref{eq.D.vac.1} and performing the $dz$ integral using the modified Dirac delta function, 
we arrive at,
\begin{eqnarray}
	\IM\Pi_D^\vac(k,p) &=& \frac{g^4}{16\pi}\frac{\partial}{\partial\lambda}\int_{0}^{1}dy
	\frac{(1-y-\ztil_+)\Theta(\ztil_+)\Theta(1-y-\ztil_+) 
		+ (1-y-\ztil_-)\Theta(\ztil_-)\Theta(1-y-\ztil_-)}{\sqrt{(M^2+sy)^2-4m^2(M^2+sy^2+\lambda)}}
	\Bigg|_{\lambda=0}~.
\end{eqnarray}
The presence of the step functions in the above equation ensure that the spikes of the Dirac delta functions were within the integration domain for a non-vanishing contribution. Finally evaluating the derivative with respect to the parameter $\lambda$, we get
\begin{eqnarray}
	\IM\Pi_D^\vac(s) = \frac{-g^4}{16\pi}\int_{0}^{1}dy\frac{(sy+M^2-2m^2)\TB{\Theta(z_+)\Theta(1-y-z_+) + \Theta(z_-)\Theta(1-y-z_-)}}{\TB{(M^2+sy)^2-4m^2(M^2+sy^2+\lambda)}^{3/2}}
	\label{eq.ImD.Final}
\end{eqnarray}
where, $z_\pm(s,y) =\ztil_\pm(s,y,\lambda=0)$ is defined in Eq.~\eqref{eq.zpm.1}. 
It can be noticed that, the $\IM\Pi_D^\vac(k,p)$ depends only on the Lorentz scalar $s=(k+p)^2$.


\section{IMAGINARY PARTS OF THE FORWARD SCATTERING AMPLITUDES IN THE WEAK FIELD APPROXIMATION} \label{appendix.impiABCD.weak}
In this appendix, we will provide the calculations of the imaginary parts of different forward scattering amplitudes under 
\textit{weak external magnetic field} as given in Eqs.~\eqref{eq.A.eB}-\eqref{eq.D.eB}.

In order to calculate $\IM\Pi_A^\text{Weak}$, we start from Eq.~\eqref{eq.ImA.2}, which in the weak field approximation becomes
\begin{eqnarray}
\IM\Pi_A^\Weak(k,p) = g^2\TB{\frac{\IM\Pi_\Weak(k+p)}{(s-M^2)^2}}~. \label{eq.ImA.3}
\end{eqnarray}
The calculation of $\IM\Pi_\Weak$ is provided in Appendix~\ref{appendix.Pi.Weak} and we get from Eqs.~\eqref{eq.ImPi.Weak} and 
\eqref{eq.Mandelstam.eB}
\begin{eqnarray}
\IM\Pi_\Weak(k+p) = \IM\Pi_\vac(k+p) +g^2\frac{(eB)^2}{24\pi}\frac{\TB{s(s-4m^2)(\sper-\spll)+12m^4\sper}\Theta(s-4m^2)}
{[s(s-4m^2)]^{5/2}}~.
\end{eqnarray}
Substituting the above equation into Eq.~\eqref{eq.ImA.3} and making use of Eqs.~\eqref{eq.ImA.2} and \eqref{eq.ImA.Final}, we finally get
\begin{eqnarray}
\IM\Pi_A^\Weak(\spll,\sper) = \IM\Pi_A^\vac(s) +
\frac{g^4}{(s-M^2)^2}\frac{(eB)^2}{24\pi}\frac{\TB{s(s-4m^2)(\sper-\spll)+12m^4\sper}\Theta(s-4m^2)}{[s(s-4m^2)]^{5/2}}~.
\label{eq.A.Weak}
\end{eqnarray}

Next we proceed to calculate $\IM\Pi_B^\text{Weak}$ from Eq.~\eqref{eq.ImB.Vac}, which in the weak field approximation becomes
\begin{eqnarray}
\IM\Pi_B^\Weak(k,p)=g\TB{\frac{-\IM\mcV_\Weak(k,p)}{(s-M^2)}}~. \label{eq.ImB.Vac.1}
\end{eqnarray}
The calculation of $\IM\mcV_\Weak$ is provided in Appendix~\ref{appendix.Vertex.Weak} and we get from Eqs.~\eqref{eq.ImVertex.Weak} and 
\eqref{eq.Mandelstam.eB}
\begin{eqnarray}
\IM\mcV_\Weak(k,p) &=& \IM\mcV_\vac(k,p) - g^3\frac{(eB)^2}{48\pi}\!\!\sum_{z\in\{\ztil_\pm\}}^{}\!\!\frac{\del^3}{\del\lambda^3}
\int_{0}^{1}\!\!dy\Theta(z)\Theta(1-y-z)z^3 \nn \\ && \hspace{0cm}
\times \TB{\frac{(2\pper^2-m^2)(1+z^2)+(2\kper^2-m^2)y^2+(\spll-\sper-2m^2+2\kper^2+2\pper^2)yz+m^2}
	{\sqrt{(M^2+sy)^2-4m^2(M^2+sy^2+\lambda)}}}\Bigg|_{\lambda=0}
\label{eq.l1}
\end{eqnarray}
where, 
\begin{eqnarray}
\ztil_\pm(s,y,\lambda) = \frac{1}{2m^2}\TB{M^2-2m^2y+sy\pm\sqrt{(M^2+sy)^2-4m^2(M^2+sy^2+\lambda)}}~. \label{eq.ztilpm.3}
\end{eqnarray}
We now substitute Eq.~\eqref{eq.l1} into Eq.~\eqref{eq.ImB.Vac.1} and perform the derivative with respect to the parameter $\lambda$. 
After some simplifications we arrive at,
\begin{eqnarray}
\IM\Pi_B^\Weak(\spll,\sper,\kper^2,\pper^2)&=&\IM\Pi_B^\vac(s)+\frac{g^4}{(s-M^2)}\frac{(eB)^2}{48\pi}\!\int_{0}^{1}\!\!dy
\TB{\Theta(z_+)\Theta(1-y-z_+)+\Theta(z_-)\Theta(1-y-z_-)} \nn \\ && \hspace{7cm} \times
\mathcal{T}_B^\Weak(\spll,\sper,\kper^2,\pper^2)
\label{eq.B.Weak}
\end{eqnarray}
where, $z_\pm=z_\pm(s,y)$ is defined in Eq.~\eqref{eq.zpm.1} and
\begin{eqnarray}
\mathcal{T}_B^\Weak(\spll,\sper,\kper^2,\pper^2) &=& \frac{1}{\TB{(M^2+sy)^2-4 m^2 (M^2+s y^2)}^{7/2}}
6\Bigg[2 m^4 \Big[M^2 y \Big\{-2 y^2 \big(\kper^2 (5 y-6)-2 (5 y-3) (\sper-\spll)\big) \nn \\ && \hspace{-3.3cm}
+2 \pper^2 (5 y^3-8y^2+15 y-6)+s (-5 y^3+3 y^2+9 y-3)\Big\} +s y^3 \Big\{2 y^2 (\kper^2+4 \sper-4 \spll) -6\pper^2 (y^2-3)-s (y^2-3)\Big\}
\nn \\ && \hspace{-3cm} +M^4 (2 y^3-9 y^2+12 y-3)\Big] 
-m^2 \Big[M^4 \Big\{4 \pper^2(8 y^3-18 y^2+14 y-3)-y \big(4 \kper^2 (6 y^2-9 y+2)+s (3 y^2+4 y-13) \nn \\ && \hspace{-3cm}
-4 (6 y^2-6 y+1) (\sper-\spll)\big)\Big\} +M^2 s y \Big\{y \big(-4 \kper^2 y (3 y-5)+s (-5 y^2+8 y+3) +8 y (3 y-2)(\sper-\spll)\big)
\nn \\ && \hspace{-3cm}
-4 \pper^2 (2 y^3+2 y^2-9 y+3)\Big\} +s^2 y^3 \Big\{4 y^2(\kper^2+\sper-\spll) +\pper^2 (12-8 y^2)+s (y^2+1)\Big\}
\nn \\ && \hspace{-3cm}
+M^6 (3 y^2-12y+11)\Big] +2 (M^2+s y) \Big[M^4 \Big\{\kper^2 (4-3 y) y +\pper^2 (6 y^2-16 y+11)+2 (y-1) y(\sper-\spll)\Big\}
\nn \\ && \hspace{-3cm}
-2 M^2 s y \Big\{\kper^2 (y-2) y +\pper^2 (2 y^2-2 y-1) -(y-1) y(\sper-\spll)\Big\}+s^2 y^2 (\kper^2 y^2+\pper^2)\Big]\nn \\ && \hspace{-3cm}
+20 m^8 (y^3+1)+2 m^6 \Big[M^2 (2y^2-15 y+6) y +y^3 \Big\{-20 \pper^2+s (2 y^2-9)+10 y^2 (\spll-\sper)\Big\}\Big]\Bigg]~.
\label{eq.TB.Weak}
\end{eqnarray}

As it turns out that the $\IM\Pi_B^\Weak(k,p)$ is a function of the Lorentz scalars $\spll=(\kpll+\ppll)^2$, 
$\sper=(\kper+\pper)^2$, $\kper^2$ and $\pper^2$, it immediately follows from Eq.~\eqref{eq.C.eB} that
\begin{eqnarray}
\IM\Pi_C^\Weak(\spll,\sper,\kper^2,\pper^2) = \IM\Pi_B^\Weak(\spll,\sper,\kper^2,\pper^2)~.
\label{eq.C.Weak}
\end{eqnarray}

Finally the calculation of $\IM\Pi_D^\text{Weak}$ is done as follows. We start with Eq.~\eqref{eq.D.eB}, which in the weak field approximation becomes
\begin{eqnarray}
\Pi_D^\Weak(k,p) &=& ig^4\!\!\int\!\!\frac{d^4\ktilde}{(2\pi)^4}
\Delta^2_F(\ktilde,M)\Delta_\Weak(k-\ktilde,m)\Delta_\Weak(p+\ktilde,m) \nn \\
&=& \Pi_D^\vac(k,p) + (eB)^2ig^4\!\!\int\!\!\frac{d^4\ktilde}{(2\pi)^4}\frac{1}{(\ktilde^2-M^2+i\epsilon)^2}
\TB{\frac{(\pper+\ktilper)^2-(\ppll+\ktilpll)^2+m^2}{\{(k-\ktilde)^2-m^2+i\epsilon\}\{(p+\ktilde)^2-m^2+i\epsilon\}^4} \right. \nn \\
	&& \left. \hspace{7cm}
	+ \frac{(\kper-\ktilper)^2-(\kpll-\ktilpll)^2+m^2}{\{(k-\ktilde)^2-m^2+i\epsilon\}^4\{(p+\ktilde)^2-m^2+i\epsilon\}}}~.
\end{eqnarray}
Using standard Feynman parametrization, we combine the denominators of the above equation and get,
\begin{eqnarray}
\Pi_D^\Weak(k,p) &=& \Pi_D^\vac(k,p) + (eB)^2 120ig^4 \!\!\int\!\!\frac{d^4\ktilde}{(2\pi)^4}\!\int_{0}^{1}\!\!\int_{0}^{1}\!\!\int_{0}^{1}\!\!dxdydz
\delta(1-x-y-z)(1-y-z)^2 \nn \\ && \hspace{1cm} \times 
\TB{\frac{z^3\SB{(\pper+\ktilper)^2-(\ppll+\ktilpll)^2+m^2}+y^3\SB{(\kper-\ktilper)^2-(\kpll-\ktilpll)^2+m^2}}
	{[(\ktilde-yk+zp)^2-\Delta_\mcV]^7}}
\end{eqnarray}
where $\Delta_\mcV= (y+z)^2m^2+(1-y-z)M^2-yzs-i\varepsilon$.
Shifting $\ktilde\rightarrow(\ktilde+yk-zp)$, the momentum integral in the 
above equation could be performed using the identities provided in Appendix~\ref{appendix.identities}. 
After some simplifications, we arrive at
\begin{eqnarray}
\Pi_D^\Weak(k,p) &=& \Pi_D^\vac(k,p) + g^4\frac{5(eB)^2}{28\pi^2}\!\int_{0}^{1}dy\!\!\int_{0}^{1-y}\!\!dz(1-y-z)^2
\Big[2y(1-y-z)\SB{y^2(1-y)-z^3}(\kper^2+\pper^2-m^2) \nn \\ && \hspace{6cm}
+yz(y^2+z^2-y^3-z^3)(\sper-\spll)+(y^3+z^3)m^2\Big]\frac{1}{\Delta_\mcV^5}~.
\end{eqnarray}
It is to be noted that, the $\Delta_\mcV$ in the above equation contains a small negative imaginary part, which will give rise to non-zero 
imaginary part of the amplitude in certain kinematic domains. 
In order to calculate the imaginary part of the $\Pi_D^\Weak$, we use the trick as given in Eq.~\eqref{eq.Trick} and write,
\begin{eqnarray}
\IM\Pi_D^\Weak(k,p) &=& \IM\Pi_D^\vac(k,p) + g^4\frac{5(eB)^2}{672\pi}\frac{\del^4}{\del\lambda^4}
\int_{0}^{1}\!\!dy\int_{0}^{1-y}\!\!dz(1-y-z)^2 \nn \\ && \hspace{-1cm}
\times \Big[2y(1-y-z)\SB{y^2(1-y)-z^3}(\kper^2+\pper^2-m^2) +yz(y^2+z^2-y^3-z^3)(\sper-\spll)+(y^3+z^3)m^2\Big]
\nn \\ && \hspace{0cm}
\times \delta\TB{(y+z)^2m^2+(1-y-z)M^2-yzs+\lambda}\Big|_{\lambda=0}~.
\label{eq.l4}
\end{eqnarray}
In order to simplify the above expression, we transform the Dirac delta function in the above equation as
\begin{eqnarray}
\delta\TB{(y+z)^2m^2+(1-y-z)M^2-yzs+\lambda} = \frac{\delta(z-\ztil_+)+\delta(z-\ztil_-)}{\sqrt{(M^2+sy)^2-4m^2(M^2+sy^2+\lambda)}}
\label{eq.delta.5}
\end{eqnarray} 
where,
\begin{eqnarray}
\ztil_\pm(s,y,\lambda) = \frac{1}{2m^2}\TB{M^2-2m^2y+sy\pm\sqrt{(M^2+sy)^2-4m^2(M^2+sy^2+\lambda)}}~. \label{eq.ztilpm.4}
\end{eqnarray}
Substituting Eq.~\eqref{eq.delta.5} into Eq.~\eqref{eq.l4} and performing the $dz$ integral using the modified Dirac delta function, we get,
\begin{eqnarray}
\IM\Pi_D^\Weak(k,p) &=& \IM\Pi_D^\vac(k,p) + g^4\frac{5(eB)^2}{672\pi}\!\!\sum_{z\in\{z_\pm\}}^{}\!\!
\frac{\del^4}{\del\lambda^4}\!\int_{0}^{1}\!\!dy\Theta(z)\Theta(1-y-z)(1-y-z)^2 \nn \\ && \hspace{-2cm}
\times \Big[\frac{2y(1-y-z)\SB{y^2(1-y)-z^3}(\kper^2+\pper^2-m^2) +yz(y^2+z^2-y^3-z^3)(\sper-\spll)+(y^3+z^3)m^2}
{\sqrt{(M^2+sy)^2-4m^2(M^2+sy^2+\lambda)}}\Big]\Big|_{\lambda=0}~.
\end{eqnarray}
Finally, performing the derivatives with respect to the parameter $\lambda$, we get after some simplifications,
\begin{eqnarray}
\IM\Pi_D^\Weak(\spll,\sper,\kper^2,\pper^2) &=& \IM\Pi_D^\vac(s) + g^4\frac{5(eB)^2}{28\pi}
\!\int_{0}^{1}\!\!dy\TB{\Theta(z_+)\Theta(1-y-z_+)+\Theta(z_-)\Theta(1-y-z_-)} \nn \\ && \hspace{7cm} \times
\mathcal{T}_D^\Weak(\spll,\sper,\kper^2,\pper^2)
\label{eq.D.Weak}
\end{eqnarray}
where, $z_\pm=z_\pm(s,y)$ is defined in Eq.~\eqref{eq.zpm.1} and
\begin{eqnarray}
\mathcal{T}_D^\Weak(\spll,\sper,\kper^2,\pper^2) &=&
\frac{1}{\TB{(M^2+s y)^2-4 m^2 (M^2+s y^2)}^{9/2}}\Bigg[-140 y^3 m^{10}+15 \{s (5-4 y) y^3+M^2 (3 y-2) y\} m^9 \nn \\ && \hspace{-3cm}
+20 y^2 \Big\{(9 y-3) M^2+y \{7 \kper^2+7 \pper^2+y(-12 y s+18 s-7 \sper+7 \spll)\} \Big\}m^8 -3 \{ (15 y^2-9 y-1) M^4\nn \\ && \hspace{-3cm}
+s y (-15 y^3+27 y^2+3 y-5) M^2+s^2y^3 (-9 y^2+4 y+10)\}m^7 +y \Big\{ (-96 y^2+72 y-6) M^4\nn \\ && \hspace{-3cm}
+3 y \{ 76 s y^3-122 s y^2+55 \sper y^2-55 \spll	y^2+16 s y+10 s -10 \sper+10 \spll+\kper^2 (20-60 y)+\pper^2 (20-60 y)\} M^2
\nn \\ && \hspace{-3cm}
-s y^3 \{ 36 s y^3-192 s y^2+80\sper y^2 -80 \spll y^2+126 s y-105 \sper y+105 \spll y+60 s-110 \sper+110 \spll \nn \\ && \hspace{-3cm}
-120 \kper^2 (2 y-3)-120\pper^2 (2 y-3)\}\Big\} m^6+\{ (15 y^2-6 y-4) M^6+3 s y (3 y^2+8 y-6) M^4 +3 \{ -24 s y^4+40 s y^3\nn \\ && \hspace{-3cm}
-3 s^2 y^2 (4 y^2-12 y+3)M^2+s^3 y^3 (-6 y^2+6 y+5)\} m^5+y \Big\{ 4 (7 y^2-9 y+2) M^6 +25 \spll y^3-8 s y^2 +2\spll y^2
\nn \\ && \hspace{-3cm}
-8 s y-13 \spll y+\spll+\pper^2 (32 y^2-24 y+2) -\sper (25 y^3+2 y^2-13 y+1)\}M^4 -3 s y \{ 18 s y^4+15 \spll y^4 -2 s y^3
\nn \\ && \hspace{-3cm}
-9 \spll y^3-30 s y^2 -54 \spll y^2+14 s y+13 \spll y+5 \spll+2\pper^2 (38 y^3-61 y^2+8 y+5) +\sper (-15 y^4+9 y^3+54 y^2
\nn \\ && \hspace{-3cm}
-13 y-5)\} M^2+s^2 y^3 \{ 6 (6 y^3-32y^2+21 y+10) \pper^2 
-2 s (y^3+13 y^2-9 y-5)+3 (\sper-\spll) (3 y^3+18 y^2 \nn \\ && \hspace{-3cm}
-26 y-10) \}+6\kper^2 \{ (16 y^2-12 y+1) M^4-s y (38 y^3-61 y^2+8 y+5) M^2 +s^2 y^3 (6 y^3-32 y^2 +21y\nn \\ && \hspace{-3cm}
+10) \} \Big\} m^4+(y-1) (M^2+s y)^3 (2 s y-3 M^2) m^3 +y \Big\{ -6 (y-1)^2 M^8+ \{ 6 s y^3-15 \spll y^3 -12s y^2-8 \spll y^2
\nn \\ && \hspace{-3cm}
+6 s y+22 \spll y-4 \spll -4 \pper^2 (7 y^2-9 y+2)+\sper (15 y^3+8 y^2-22y+4) \} M^6+3 s y \{ 8 s y^3+9 \spll y^3-16 s y^2
\nn \\ && \hspace{-3cm}
-28 \spll y^2+8 s y+10 \spll y+4 \spll+8 \pper^2(y-1)^2 (3 y+1) -\sper (9 y^3-28 y^2+10 y+4) \} M^4 +3 s^2 y^2 \{ 2 s y^3
\nn \\ && \hspace{-3cm}
+13 \spll y^3-4 s y^2-13 \spll y^2+2 sy-12 \spll y 
+7 \spll+\sper (-13 y^3+13 y^2+12 y-7)+2 \pper^2 (9 y^3-y^2-15 y+7) \} M^2 \nn \\ && \hspace{-3cm}
+2\kper^2 (y-1) \{ (4-14 y) M^6+12 s y (3 y^2-2 y-1) M^4+3 s^2 y^2 (9 y^2+8 y-7) M^2 +s^3 y^3 (y^2+14y+5) \}
\nn \\ && \hspace{-3cm}
+s^3 y^3 \{ -6 s y^3+7 \spll y^3+12 s y^2+7 \spll y^2-6 s y-14 \spll y-5 \spll +\sper (-7y^3-7 y^2+14 y+5)\nn \\ && \hspace{-3cm}
+2 \pper^2 (y^3+13 y^2-9 y-5) \} \Big\} m^2 
+(y-1) y (M^2+s y)^2 \Big\{ 3 \{ 2\pper^2 (y-1)-(\sper-\spll) (2 y-1) \} M^4 \nn \\ && \hspace{-3cm}
+s y \{ (\sper-\spll) (8 y-9)
-18 \pper^2 (y-1) \} M^2+s^2\{ 6 (y-1) \pper^2+(\spll-\sper) (y-3)\} y^2 \nn \\ && \hspace{-3cm}
+6 \kper^2 (y-1) (M^4-3 s y M^2+s^2y^2) \Big\}\Bigg]~.
\label{eq.TD.Weak}
\end{eqnarray}
The presence of the step functions in the above equation ensure that the spikes of the Dirac delta functions 
were within the integration domain for a non-vanishing contribution.


\section{IMAGINARY PARTS OF THE FORWARD SCATTERING AMPLITUDES IN THE STRONG FIELD APPROXIMATION} \label{appendix.impiABCD.strong}
In this appendix, we will provide the calculations of the imaginary parts of different forward scattering amplitudes under 
\textit{strong external magnetic field} as given in Eqs.~\eqref{eq.A.eB}-\eqref{eq.D.eB}.

For the evaluation of $\IM\Pi_A^\text{Strong}$, we start with Eq.~\eqref{eq.ImA.2}, which in the strong field approximation becomes
\begin{eqnarray}
\IM\Pi_A^\Strong(k,p) = g^2\TB{\frac{\IM\Pi_\Strong(k+p)}{(s-M^2)^2}}~. \label{eq.ImA.4}
\end{eqnarray}
The calculation of $\IM\Pi_\Strong$ is provided in Appendix~\ref{appendix.Pi.Strong} and we get from Eqs.~\eqref{eq.ImPi.Strong} and 
\eqref{eq.Mandelstam.eB}
\begin{eqnarray}
\IM\Pi_\Strong(q) = -g^2\frac{eB}{4\pi}\exp\FB{\frac{\sper}{2eB}}\frac{\Theta(\spll-4m^2-4eB)}{\sqrt{\spll(\spll-4m^2-4eB)}}~.
\end{eqnarray}
Substituting the above equation into Eq.~\eqref{eq.ImA.4}, we finally get
\begin{eqnarray}
\IM\Pi_A^\Strong(\spll,\sper) = 
\frac{-g^4}{(s-M^2)^2}\frac{eB}{4\pi}\exp\FB{\frac{\sper}{2eB}}\frac{\Theta(\spll-4m^2-4eB)}{\sqrt{\spll(\spll-4m^2-4eB)}}~.
\label{eq.A.Strong}
\end{eqnarray}

Next, for the calculation of $\IM\Pi_B^\text{Strong}$, we start with Eq.~\eqref{eq.ImB.Vac}, which in the strong field approximation becomes
\begin{eqnarray}
\IM\Pi_B^\Strong(k,p)=g\TB{\frac{-\IM\mcV_\Strong(k,p)}{(s-M^2)}}~. \label{eq.ImB.Vac.2}
\end{eqnarray}
The calculation of $\IM\mcV_\Strong$ is provided in Appendix~\ref{appendix.Vertex.Strong} 
and we get from Eqs.~\eqref{eq.ImVertex.Weak} and \eqref{eq.Mandelstam.eB}
\begin{eqnarray}
\IM\mcV_\Strong(k,p) &=& g^3\frac{eB}{8\pi}\exp\TB{\frac{\kper^2+\pper^2}{eB}}
\int_{0}^{\infty}\!\!d\xi e^{-\xi}I_0\FB{\sqrt{\frac{\xi}{2eB}\SB{\frac{}{}\sper-2\kper^2-2\pper^2}}} 
\nn \\ && \hspace{1cm} \times
\frac{\del}{\del\lambda}\int_{0}^{1}\!\!dy 
\frac{\Theta(\Ztil_+)\Theta(1-y-\Ztil_+)+\Theta(\Ztil_-)\Theta(1-y-\Ztil_-)}{\sqrt{\mathcal{D}}}
\Bigg|_{\lambda=0} \label{eq.l3}
\end{eqnarray}
where, $\Ztil_\pm$ can be obtained from Eq.~\eqref{eq.Zpm} using Eq.~\eqref{eq.Mandelstam.eB} as
\begin{eqnarray}
\Ztil_\pm(y,\lambda) = \frac{1}{4 (m^2-\pper^2)}\TB{eB(\xi-2)+2y(\spll-2m^2+\kper^2+\pper^2)+2(M^2-\pper^2) \pm 2 \sqrt{\mathcal{D}}}
\label{eq.Z}
\end{eqnarray}
with 
\begin{eqnarray}
\mathcal{D} &=& \TB{eB\FB{\frac{\xi}{2}-1}+y(\spll-2m^2+\kper^2+\pper^2)+(M^2-\pper^2)}^2 \nn \\ &&
-4 (m^2-\pper^2) \TB{eB\FB{\frac{\xi}{2}(1-y)+y}-(y^2-y)\kper^2+m^2y^2+M^2(1-y)+\lambda}~.
\end{eqnarray}
We now substitute Eq.~\eqref{eq.l3} into Eq.~\eqref{eq.ImB.Vac.2} and perform the derivatives with respect to the parameter $\lambda$. 
After some simplifications we arrive at,
\begin{eqnarray}
\IM\Pi_B^\Strong(\spll,\sper,\kper^2,\pper^2) &=& -\frac{g^4}{(s-M^2)}\frac{eB}{8\pi}\exp\TB{\frac{\kper^2+\pper^2}{eB}}
\int_{0}^{\infty}\!\!d\xi e^{-\xi}I_0\FB{\sqrt{\frac{\xi}{2eB}\SB{\frac{}{}\sper-2\kper^2-2\pper^2}}} 
\nn \\ && \hspace{0cm} \times
\int_{0}^{1}\!\!dy\TB{\Theta(Z_+)\Theta(1-y-Z_+)+\Theta(Z_-)\Theta(1-y-Z_-)}\mathcal{T}_B^\Strong(\spll,\kper^2,\pper^2)
\label{eq.B.Strong}
\end{eqnarray}
where, $Z_\pm = \Ztil_\pm(\lambda=0)$ and
\begin{eqnarray}
\mathcal{T}_B^\Strong(\spll,\kper^2,\pper^2) &=& 
2 (m^2-\pper^2)\Big[\frac{1}{4} \Big\{ eB (\xi -2)+2 \{ y (\kper^2-2m^2+\spll)+M^2+\pper^2 (y-1) \} \Big\}^2 \nn \\ &&
-4 (m^2-\pper^2)\Big\{ y \{ eB+\kper^2(1-y)+m^2 y\}-\frac{1}{2} eB \xi  (y-1)+M^2 (1-y) \Big\}\Big]^{-3/2}~.
\label{eq.TB.Strong}
\end{eqnarray}

As it turns out that the $\IM\Pi_B^\Strong(k,p)$ is a function of the Lorentz scalars $\spll=(\kpll+\ppll)^2$, 
$\sper=(\kper+\pper)^2$, $\kper^2$ and $\pper^2$, it immediately follows from Eq.~\eqref{eq.C.eB} that
\begin{eqnarray}
\IM\Pi_C^\Strong(\spll,\sper,\kper^2,\pper^2) = \IM\Pi_B^\Strong(\spll,\sper,\kper^2,\pper^2)~.
\label{eq.C.Strong}
\end{eqnarray}

Finally, for the calculation of $\IM\Pi_D^\text{Strong}$, we start with Eq.~\eqref{eq.D.eB}, 
which in the strong field approximation becomes
\begin{eqnarray}
\Pi_D^\Strong(k,p) &=& ig^4\!\!\int\!\!\frac{d^4\ktilde}{(2\pi)^4}
\Delta^2_F(\ktilde,M)\Delta_\Strong(k-\ktilde,m)\Delta_\Strong(p+\ktilde,m) \nn \\
&=& ig^4\!\!\int\!\!\frac{d^4\ktilde}{(2\pi)^4}\exp\TB{\frac{(\kper-\ktilper)^2+(\pper+\ktilper)^2}{eB}} \nn \\ && \hspace{2cm}\times
\TB{\frac{1}{(\ktilde^2-M^2+i\epsilon)^2\{(\kpll-\ktilpll)^2-m^2-eB+i\epsilon\}\{(\ppll+\ktilpll)^2-m^2-eB+i\epsilon\}}}~.
\end{eqnarray}
Using standard Feynman parametrization, we combine the denominator of the above equation and get,
\begin{eqnarray}
\Pi_D^\Strong(k,p) &=& 24ig^4\!\!\int\!\!\frac{d^2\ktilper}{(2\pi)^2}\exp\TB{\frac{(\kper-\ktilper)^2+(\pper+\ktilper)^2}{eB}} 
\nn \\ && \hspace{1cm} \times
\int_{0}^{1}\!\!\int_{0}^{1}\!\!\int_{0}^{1}\!\!dxdydz\delta(1-x-y-z) \!\int\!\!\frac{d^2\ktilpll}{(2\pi)^2}
\frac{(1-y-z)}{[(\ktilpll-y\kpll+z\ppll)^2-\tilde{\Delta}_{\mcV}]^4}
\end{eqnarray}
where,
\begin{eqnarray}
\tilde{\Delta}_{\mcV} = (y^2-y)\kpll^2+(z^2-z)\ppll^2-2yz(\kpll\cdot\ppll)+(y+z)(m^2+eB)+(1-y-z)(M^2-\ktilper^2)-i\epsilon~.
\end{eqnarray}
Shifting $\ktilpll\rightarrow(\ktilpll+y\kpll-z\ppll)$, the $d^2\ktilpll$ integral in the 
above equation could be performed using the identities provided in Appendix~\ref{appendix.identities}. After some simplifications, 
we arrive at
\begin{eqnarray}
\Pi_D^\Strong(k,p) &=& \frac{-2g^4}{\pi}\!\int_{0}^{1}\!\!dy\int_{0}^{1-y}\!\!dz(1-y-z)
\!\int\!\!\frac{d^2\ktilper}{(2\pi)^2}\exp\TB{\frac{(\kper-\ktilper)^2+(\pper+\ktilper)^2}{eB}}\frac{1}{\tilde{\Delta}_\mcV^3}~.
\end{eqnarray}
Next we go to the cylindrical polar coordinate and make the substitution $\xi=-2\kper^2/eB$ which implies 
$d^2\ktilper=\frac{eB}{4}\kperv d\kperv d\phi$. The integration over the azimuthal angle $d\phi$ could then be analytically performed 
and be expressed in terms of the modified Bessel function of first kind $I_0$ (see Appendix~\ref{appendix.identities}). We finally get
\begin{eqnarray}
\Pi_D^\Strong(k,p) &=& -g^4\frac{eB}{4\pi^2}\exp\TB{\frac{\kper^2+\pper^2}{eB}}
\int_{0}^{\infty}\!\!d\xi e^{-\xi}I_0\FB{\sqrt{\frac{\xi}{2eB}\SB{\frac{}{}\sper-2\kper^2-2\pper^2}}} 
\int_{0}^{1}\!\!dy\int_{0}^{1-y}\!\!dz\frac{1}{\tilde{\Delta}_\mcV^3}~.
\end{eqnarray}
It is to be noted that, the $\tilde{\Delta}_\mcV$ in the above equation contains a small negative imaginary part, 
which will give rise to non-zero imaginary part of the amplitude in certain kinematic domains. 
In order to calculate the imaginary part of $\Pi_D^\Strong$, we use the trick as given in Eq.~\eqref{eq.Trick} and write,
\begin{eqnarray}
\IM\Pi_D^\Strong(k,p) &=& -g^4\frac{eB}{8\pi}\exp\TB{\frac{\kper^2+\pper^2}{eB}}
\int_{0}^{\infty}\!\!d\xi e^{-\xi}I_0\FB{\sqrt{\frac{\xi}{2eB}\SB{\frac{}{}\sper-2\kper^2-2\pper^2}}} 
\frac{\del^2}{\del\lambda^2}\int_{0}^{1}\!\!dy\int_{0}^{1-y}\!\!dz(1-y-z) \nn \\ && \hspace{-1cm} \times
\delta\TB{(y^2-y)\kpll^2+(z^2-z)\ppll^2-2yz(\kpll\cdot\ppll)+(y+z)(m^2+eB)+(1-y-z)(M^2+eB\xi/2)+\lambda}\Bigg|_{\lambda=0}~.
\label{eq.l5}
\end{eqnarray}
In order to simplify, we transform the Dirac delta function in the above equation as
\begin{eqnarray}
&& \delta\TB{(y^2-y)\kpll^2+(z^2-z)\ppll^2-2yz(\kpll\cdot\ppll) + (y+z)(m^2+eB)+(1-y-z)(M^2+eB\xi/2)+\lambda} \nn \\ && \hspace{2cm}
= \frac{1}{\sqrt{\mathcal{D}}}\TB{\delta(z-\Ztil_+)+\delta(z-\Ztil_-)}
\label{eq.delta.6}
\end{eqnarray} 
where, 
\begin{eqnarray}
\Ztil_\pm = \frac{1}{4 \ppll^2}\TB{eB \xi -2 eB+4y\kpll\cdot\ppll-2 m^2+2 M^2+2\ppll^2 \pm 2 \sqrt{\mathcal{D}}}
\label{eq.Zpm.1}
\end{eqnarray}
with 
\begin{eqnarray}
\mathcal{D} &=& \TB{\frac{1}{2} eB (\xi -2)+2y\kpll\cdot\ppll -m^2+M^2+\ppll^2}^2 \nn \\ &&
-4 \ppll^2 \TB{\frac{1}{2} eB \{\xi (1-y) +2 y\} +\kpll^2 y^2-\kpll^2 y+m^2 y-M^2 (y-1)+\lambda}~.
\end{eqnarray}
Substituting Eq.~\eqref{eq.delta.6} into Eq.~\eqref{eq.l5} and performing the $dz$ integral using the modified Dirac delta function, we get,
\begin{eqnarray}
\IM\Pi_D^\Strong(k,p) &=& -g^4\frac{eB}{8\pi}\exp\TB{\frac{\kper^2+\pper^2}{eB}}
\int_{0}^{\infty}\!\!d\xi e^{-\xi}I_0\FB{\sqrt{\frac{\xi}{2eB}\SB{\frac{}{}\sper-2\kper^2-2\pper^2}}} 
\nn \\ && \hspace{1cm} \times
\frac{\del^2}{\del\lambda^2}\int_{0}^{1}\!\!dy \!\!\sum_{z\in\{\Ztil_\pm\}}^{}\!\!\!
\frac{(1-y-z)\Theta(z)\Theta(1-y-z)}{\sqrt{\mathcal{D}}}
\Bigg|_{\lambda=0}~. \label{eq.ImVertex.Strong.1}
\end{eqnarray}
The presence of the step functions in the above equation ensure that the spikes of the Dirac delta functions 
were within the integration domain for a non-vanishing contribution. 
Perform the derivatives with respect to the parameter $\lambda$, we get after some simplifications,
\begin{eqnarray}
\IM\Pi_D^\Strong(\spll,\sper,\kper^2,\pper^2) &=& -g^4\frac{eB}{8\pi}\exp\TB{\frac{\kper^2+\pper^2}{eB}}
\int_{0}^{\infty}\!\!d\xi e^{-\xi}I_0\FB{\sqrt{\frac{\xi}{2eB}\SB{\frac{}{}\sper-2\kper^2-2\pper^2}}} 
\nn \\ && \hspace{0cm} \times
\int_{0}^{1}\!\!dy\TB{\Theta(Z_+)\Theta(1-y-Z_+)+\Theta(Z_-)\Theta(1-y-Z_-)}\mathcal{T}_D^\Strong(\spll,\kper^2,\pper^2)
\label{eq.D.Strong}
\end{eqnarray}
where, $Z_\pm = \Ztil_\pm(\lambda=0)$ and
\begin{eqnarray}
\mathcal{T}_D^\Strong(\spll,\kper^2,\pper^2) &=& 
-\Big[96 (m^2-\pper^2) \Big\{ eB (\xi -2)+2 (\kper^2 y-2 m^2+M^2-\pper^2 y+\pper^2+\spll y)\Big\}\Big] \nn \\ && \times
\Big[eB^2 (\xi -2)^2+4 eB (\kper^2 \xi  y-2 \kper^2 y-2 m^2 \xi +M^2 (\xi -2)+\pper^2 \{\xi(1-y)+2 y+2\}+\xi\spll y-2 \spll y) \nn \\ &&
+4 \Big\{ \kper^4 y^2-4 m^2 \{ y (\kper^2-\pper^2+\spll y)+M^2 \}+2 M^2 \{y (\kper^2+\spll)-\pper^2 (y-1)\}-2 \kper^2 \pper^2 y^2 \nn \\ &&
+2\kper^2 \pper^2 y+2 \kper^2 \spll y^2+M^4+\pper^4 y^2-2 \pper^4 y+\pper^4+2 \pper^2 \spll	y^2-2 \pper^2 \spll y
+\spll^2 y^2\Big\}\Big]^{-5/2}~.
\label{eq.TD.Strong}
\end{eqnarray}


\section{ONE-LOOP SELF-ENERGY OF $B^0$} 
In this appendix, we will provide the calculation of one-loop self-energy of $B^0$. First we will obtain the expression of 
the vacuum self-energy (i.e in absence of external magnetic field), followed the evaluation of the same under external magnetic 
field employing both the weak and strong magnetic field approximations. We do these in the following three subsections:

\subsection{VACUUM SELF-ENERGY} \label{appendix.Pi.Vac}
 the calculation of the imaginary part of one-loop \textit{vacuum} self-energy of $B^0$ due to 
$b^+b^-$ loop. We start with Eq.~\eqref{eq.Pi.Vac}
\begin{eqnarray}
\Pi_\vac(q) = ig^2\!\!\int\!\!\frac{d^4\ktilde}{(2\pi)^4}\Delta_F(\ktilde,m)\Delta_F(q+\ktilde,m) 
= ig^2\!\int\!\frac{d^4\ktilde}{(2\pi)^4}\frac{1}{(\ktilde^2-m^2+i\epsilon)\{(q+\ktilde)^2-m^2+i\epsilon\}}~.
\end{eqnarray}
Using standard Feynman parametrization, we combine the denominator of the above equation and get,
\begin{eqnarray}
\Pi_\vac(q) = ig^2\!\!\int\!\!\frac{d^4\ktilde}{(2\pi)^4}\int_{0}^{1}\!\!dx\frac{1}{[(\ktilde+xq)^2-\Delta_\Pi]^2}
\end{eqnarray}
where, $\Delta_\Pi=m^2-x(1-x)q^2-i\epsilon$. Shifting $\ktilde\rightarrow(\ktilde-xq)$, the momentum integral in the 
above equation could be performed using the identities 
provided in Appendix~\ref{appendix.identities} and we arrive at
\begin{eqnarray}
\Pi_\vac(q) = \frac{-g^2}{16\pi^2}\int_{0}^{1}\!\!dx\Gamma(\varepsilon)\FB{\frac{\mu}{\Delta_\Pi}}^\varepsilon\Bigg|_{\varepsilon\rightarrow0}
\end{eqnarray}
where $\varepsilon=(2-d/2)$. Here, the space-time dimension has been changed from $4$ to $d$ following the dimensional regularization, so 
that the scale parameter $\mu$ with dimension GeV$^2$ has been introduced to keep the overall dimension of the self-energy same. 
Expanding the above equation about $\varepsilon=0$, we get
\begin{eqnarray}
\Pi_\vac(q) = \frac{-g^2}{16\pi^2}\int_{0}^{1}\!\!dx
\TB{\frac{1}{\varepsilon}-\gamma_\text{E}-\ln\FB{\frac{\Delta_\Pi}{\mu}}}\Bigg|_{\varepsilon\rightarrow0}
\end{eqnarray}
where, $\gamma_\text{E}$ is the Euler-Mascheroni constant. 
It is to be noted that, the $\Delta_\Pi$ in the above equation contains a small negative imaginary part, which will give rise to non-zero 
imaginary part of self-energy in certain kinematic domains. The imaginary part of the self-energy follows from the branch cut of the 
logarithm as
\begin{eqnarray}
\IM\Pi_\vac(q) = \frac{g^2}{16\pi^2}\int_{0}^{1}\!\!dx(-\pi)\Theta\TB{-m^2+x(1-x)q^2}
\end{eqnarray}
where $\Theta(x)$ is the unit step function. The presence of the step function in the above equation will restrict the limits 
of $dx$ integration and we get
\begin{eqnarray}
\IM\Pi_\vac(q) = -\frac{g^2}{16\pi}\int_{x_-}^{x_+}\!\!dx\Theta(q^2-4m^2)
\end{eqnarray}
where $x_\pm = \FB{\frac{1}{2}\pm\sqrt{\frac{1}{4}-\frac{m^2}{q^2}}}$. Performing the $dx$ integral in the above equation we finally get,
\begin{eqnarray}
\IM\Pi_\vac(q) = -\frac{g^2}{16\pi\sqrt{q^2}}\sqrt{q^2-4m^2}\Theta(q^2-4m^2)~. \label{eq.ImPi.Vac}
\end{eqnarray}


\subsection{SELF-ENERGY IN THE WEAK FIELD APPROXIMATION} \label{appendix.Pi.Weak}
In this appendix, we will show the calculation of the imaginary part of the one-loop self-energy of $B^0$ under external 
magnetic field by employing the weak field approximation.
We start with Eq.~\eqref{eq.Pi.eB}
\begin{eqnarray}
\Pi_\Weak(q) &=& ig^2\!\!\int\!\!\frac{d^4\ktilde}{(2\pi)^4}\Delta_\Weak(\ktilde,m)\Delta_\Weak(q+\ktilde,m)
\end{eqnarray}
and substitute the weak field expansion of the Schwinger propagator from Eq.~\eqref{eq.Schwinger.Weak} in the above equation to get,
\begin{eqnarray}
\Pi_\Weak(q) &=& \Pi_\vac(q) + (eB)^2 2ig^2\!\!\int\!\!\frac{d^4\ktilde}{(2\pi)^4}\frac{(\ktilper^2-\ktilpll^2+m^2)}{(\ktilde^2-m^2+i\epsilon)^4\{(q+\ktilde)^2-m^2+i\epsilon\}}~.
\end{eqnarray}
Using standard Feynman parametrization, we combine the denominator of the above equation and get,
\begin{eqnarray}
\Pi_\Weak(q) = \Pi_\vac(q) +(eB)^2 8ig^2\!\!\int\!\!\frac{d^4\ktilde}{(2\pi)^4}\int_{0}^{1}\!\!dx(1-x)^3\frac{(\ktilper^2-\ktilpll^2+m^2)}{[(\ktilde+xq)^2-\Delta_\Pi]^5}
\end{eqnarray}
where, $\Delta_\Pi=m^2-x(1-x)q^2-i\epsilon$. Shifting $\ktilde\rightarrow(\ktilde-xq)$, the momentum integral in the 
above equation could be performed using the identities given in Appendix~\ref{appendix.identities} and we arrive at
\begin{eqnarray}
\Pi_\Weak(q) = \Pi_\vac(q) +(eB)^2\frac{g^2}{24\pi^2}\int_{0}^{1}\!\!dx(1-x)^3\frac{m^2+x^2(\qper^2-\qpll^2)}{\Delta_\Pi^3}~.
\end{eqnarray}
It is to be noted that, the $\Delta_\Pi$ in the above equation contains a small negative imaginary part, which will give rise to non-zero 
imaginary part of self-energy in certain kinematic domains. 
In order to calculate the imaginary part of the self-energy, we use the trick as given in Eq.~\eqref{eq.Trick} and write,
\begin{eqnarray}
\IM\Pi_\Weak(q) = \IM\Pi_\vac(q) +(eB)^2\frac{g^2}{48\pi}\frac{\del^2}{\del\lambda^2}
\int_{0}^{1}\!\!dx(1-x)^3\SB{m^2+x^2(\qper^2-\qpll^2)}\delta\TB{m^2-x(1-x)q^2+\lambda}\Bigg|_{\lambda=0}~. \label{eq.Pi.Weak.1}
\end{eqnarray}
We now transform the Dirac delta function as
\begin{eqnarray}
\delta\TB{m^2-x(1-x)q^2+\lambda} = \frac{\delta(x-x_+)+\delta(x-x_-)}{\sqrt{q^2(q^2-4m^2-4\lambda)}}
\end{eqnarray}
where, $x_\pm = \frac{1}{2}\pm\frac{1}{2q^2}\sqrt{q^2(q^2-4m^2-4\lambda)}$. We now substitute the above equation 
into Eq.~\eqref{eq.Pi.Weak.1} and perform the integral over $dx$ using the Dirac delta functions to get,
\begin{eqnarray}
\IM\Pi_\Weak(q) = \IM\Pi_\vac(q) +(eB)^2\frac{g^2}{48\pi}\frac{\del^2}{\del\lambda^2}\!\!\sum_{x\in\{x_\pm\}}^{}\!
\TB{\frac{(1-x)^3\{m^2+x^2(\qper^2-\qpll^2)\}\Theta(x)\Theta(1-x)}{\sqrt{q^2(q^2-4m^2-4\lambda)}}}\Bigg|_{\lambda=0}~. 
\end{eqnarray}
The presence of the step functions in the above equation ensure that the spikes of the Dirac delta functions 
were within the integration domain for a non-vanishing contribution. 
Evaluating the derivative with respect to the parameter $\lambda$, we get after some simplification
\begin{eqnarray}
\IM\Pi_\Weak(q) = \IM\Pi_\vac(q) +g^2\frac{(eB)^2}{24\pi}\frac{\TB{q^2(q^2-4m^2)(\qper^2-\qpll^2)+12m^4\qper^2}\Theta(q^2-4m^2)}
{[q^2(q^2-4m^2)]^{5/2}}~.
\label{eq.ImPi.Weak}
\end{eqnarray}

\subsection{SELF-ENERGY IN THE STRONG FIELD APPROXIMATION} \label{appendix.Pi.Strong}
In this appendix, we will show the calculation of the imaginary part of the one-loop self-energy of $B^0$ under external 
magnetic field by employing the strong field approximation.
We start with Eq.~\eqref{eq.Pi.eB}
\begin{eqnarray}
\Pi_\Strong(q) &=& ig^2\!\!\int\!\!\frac{d^4\ktilde}{(2\pi)^4}\Delta_\Strong(\ktilde,m)\Delta_\Strong(q+\ktilde,m)
\end{eqnarray}
and substitute the strong field Schwinger propagator from Eq.~\eqref{eq.Schwinger.LLL} in the above equation  to obtain,
\begin{eqnarray}
\Pi_\Strong(q) = 4ig^2\!\!\int\!\!\frac{d^4\ktilde}{(2\pi)^4}\exp\TB{\frac{-\ktilper^2-(\qper+\ktilper)^2}{eB}}
\frac{1}{(\ktilpll^2-m^2-eB+i\epsilon)\{(\qpll+\ktilpll)^2-m^2-eB+i\epsilon\}}~.
\end{eqnarray}
Using standard Feynman parametrization, we combine the denominator of the above equation and get,
\begin{eqnarray}
\Pi_\Strong(q) = 4ig^2\!\!\int\!\!\frac{d^2\ktilper}{(2\pi)^2}\exp\TB{\frac{-\ktilper^2-(\qper+\ktilper)^2}{eB}}
\int\!\!\frac{d^2\ktilpll}{(2\pi)^2}\int_{0}^{1}\!\!dx
\frac{1}{[(\ktilpll+x\qpll)^2-\Delta_\parallel]^2}
\end{eqnarray}
where, $\Delta_\parallel=m^2+eB-x(1-x)\qpll^2-i\epsilon$. Shifting $\ktilpll\rightarrow(\ktilpll-x\qpll)$, the $d^2\ktilpll$ integral in the 
above equation could be performed using the identities provided in Appendix~\ref{appendix.identities}. Also performing 
the remaining Gausian integral over $d^2\ktilper$ we get, 
\begin{eqnarray}
\Pi_\Strong(q) = -g^2\frac{eB}{8\pi^2}\exp\TB{\frac{\qper^2}{2eB}}\int_{0}^{1}\!\!dx\frac{1}{\Delta_\parallel}~.
\end{eqnarray}
It is to be noted that, the $\Delta_\parallel$ in the above equation contains a small negative imaginary part, 
which will give rise to non-zero imaginary part of self-energy in certain kinematic domains. 
Taking the imaginary part of the above equation, we get
\begin{eqnarray}
\IM\Pi_\Strong(q) = -g^2\frac{eB}{8\pi}\exp\TB{\frac{\qper^2}{2eB}}\int_{0}^{1}\!\!dx\delta\TB{m^2+eB-x(1-x)\qpll^2}~.
\label{eq.Pi.Strong.1}
\end{eqnarray}
We now transform the Dirac delta function as
\begin{eqnarray}
\delta\TB{m^2+eB-x(1-x)\qpll^2} = \frac{\delta(x-X_+)+\delta(x-X_-)}{\sqrt{\qpll^2(\qpll^2-4m^2-4eB)}}
\end{eqnarray}
where, $X_\pm = \frac{1}{2}\pm\frac{1}{2\qpll^2}\sqrt{\qpll^2(\qpll^2-4m^2-4eB)}$. We now substitute the above equation 
into Eq.~\eqref{eq.Pi.Strong.1} and perform the integral over $dx$ using the Dirac delta functions. After some simplifications, we 
obtain,
\begin{eqnarray}
\IM\Pi_\Strong(q) = -g^2\frac{eB}{4\pi}\exp\FB{\frac{\qper^2}{2eB}}\frac{\Theta(\qpll^2-4m^2-4eB)}{\sqrt{\qpll^2(\qpll^2-4m^2-4eB)}}~.
\label{eq.ImPi.Strong}
\end{eqnarray}
The presence of the step functions in the above equation ensure that the spikes of the Dirac delta functions 
were within the integration domain for a non-vanishing contribution.


\section{ONE-LOOP $B^0b^+b^-$ VERTEX FUNCTION} 
In this appendix, we will provide the calculation of one-loop $B^0b^+b^-$ vertex function. First we will obtain the expression of 
the vacuum vertex function (i.e in absence of external magnetic field), followed the evaluation of the same under external magnetic 
field employing both the weak and strong magnetic field approximations. We do these in the following three subsections:

\subsection{VERTEX FUNCTION IN THE VACUUM }\label{appendix.Vertex.Vac}
In this appendix, we will calculate of the imaginary part of the one-loop vacuum vertex function. 
We have from Eq.~\eqref{eq.Vertex.Vac}
\begin{eqnarray}
\mcV_\vac(k,p) &=& ig^3\!\!\int\!\!\frac{d^4\ktilde}{(2\pi)^4}\Delta_F(\ktilde,M)\Delta_F(k-\ktilde,m)\Delta_F(p+\ktilde,m) \nn \\
&=& -ig^3\!\!\int\!\!\frac{d^4\ktilde}{(2\pi)^4}\frac{1}{(\ktilde^2-M^2+i\epsilon)\{(k-\ktilde)^2-m^2+i\epsilon\}\{(p+\ktilde)^2-m^2+i\epsilon\}}~.
\end{eqnarray}
Using Feynman parametrization, we combine the denominator of the above equation and get,
\begin{eqnarray}
\mcV_\vac(k,p) &=& -2ig^3\!\!\int\!\!\frac{d^4\ktilde}{(2\pi)^4}\int_{0}^{1}\!\!\int_{0}^{1}\!\!\int_{0}^{1}\!\!dxdydz
\delta(1-x-y-z)\frac{1}{[(\ktilde-yk+zp)^2-\Delta_\mcV]^3} \label{eq.Vertex.Vac.1}
\end{eqnarray}
where
\begin{eqnarray}
\Delta_\mcV = (yk-zp)^2-yk^2-zp^2+(1-y-z)M^2+(y+z)m^2-i\varepsilon~.
 \label{eq.Delta.V.Vac}
\end{eqnarray}
Next we shift $\ktilde\rightarrow(\ktilde+yk-zp)$ in Eq.~\eqref{eq.Vertex.Vac.1} and perform the momentum integration using the 
identities provided in Appendix~\ref{appendix.identities} to obtain
\begin{eqnarray}
\mcV_\vac(k,p) &=& \frac{-g^3}{16\pi^2}\int_{0}^{1}\!\!\int_{0}^{1}\!\!\int_{0}^{1}\!\!dxdydz
\delta(1-x-y-z)\frac{1}{\Delta_\mcV}~.
\end{eqnarray}
The $dx$ integral in the above equation is performed using the Dirac delta function present in the integrand and we get,
\begin{eqnarray}
\mcV_\vac(k,p) &=& \frac{-g^3}{16\pi^2}\int_{0}^{1}\!\!dy\int_{0}^{1-y}\!\!dz\frac{1}{\Delta_\mcV} \label{eq.Vertex.Vac.2}~.
\end{eqnarray}
The expression of $\Delta_\mcV$ given in Eq.~\eqref{eq.Delta.V.Vac} can be simplified by putting on-shell conditions $k^2=p^2=m^2$ 
as
\begin{eqnarray}
\Delta_\mcV = (y+z)^2m^2+(1-y-z)M^2-yz(k+p)^2-i\varepsilon \label{eq.Delta.V.Vac.2}~.
\end{eqnarray}
The presence of small negative imaginary part in $\Delta_\mcV$ will give rise to the imaginary part of the vertex function in 
certain kinematic domains. Taking imaginary part of Eq.~\eqref{eq.Vertex.Vac.2}, we get
\begin{eqnarray}
\IM\mcV_\vac(k,p) &=& \frac{-g^3}{16\pi}\int_{0}^{1}\!\!dy\int_{0}^{1-y}\!\!dz\delta\TB{(y+z)^2m^2+(1-y-z)M^2-yz(k+p)^2}
\label{eq.Im.Vertex.Vac}~.
\end{eqnarray}
In order to simplify, we transform the Dirac delta function in the above equation as
\begin{eqnarray}
\delta\TB{(y+z)^2m^2+(1-y-z)M^2-yz(k+p)^2} = \frac{\delta(z-z_+)+\delta(z-z_-)}{\sqrt{(M^2+q^2y)^2-4m^2(M^2+q^2y^2)}}
\label{eq.delta.1}
\end{eqnarray} 
where $q=(k+p)$ and 
\begin{eqnarray}
z_\pm = \frac{1}{2m^2}\TB{M^2-2m^2y+q^2y\pm\sqrt{(M^2+q^2y)^2-4m^2(M^2+q^2y^2)}}~. \label{eq.zpm}
\end{eqnarray}
Substituting Eq.~\eqref{eq.delta.1} into Eq.~\eqref{eq.Im.Vertex.Vac} and performing the $dz$ integral using the modified 
Dirac delta function, we get,
\begin{eqnarray}
\IM\mcV_\vac(k,p) &=& \frac{-g^3}{16\pi}\int_{0}^{1}\!\!dy\frac{\Theta(z_+)\Theta(1-y-z_+)+\Theta(z_-)\Theta(1-y-z_-)}
{\sqrt{(M^2+q^2y)^2-4m^2(M^2+q^2y^2)}}~. \label{eq.Im.Vertex.Vac.1}
\end{eqnarray}
The presence of the step functions in the above equation ensure that the spikes of the Dirac delta functions were within the integration domain for a non-vanishing contribution.

\subsection{VERTEX FUNCTION IN THE WEAK FIELD APPROXIMATION} \label{appendix.Vertex.Weak}
In this appendix, we will briefly sketch the calculation of the imaginary part of one-loop $B^0b^+b^-$ vertex function 
under external magnetic field by employing the weak field approximation.
We have from Eq.~\eqref{eq.Vertex.eB}
\begin{eqnarray}
\mcV_\Weak(k,p) &=& ig^3\!\!\int\!\!\frac{d^4\ktilde}{(2\pi)^4}\Delta_F(\ktilde,M)\Delta_\Weak(k-\ktilde,m)\Delta_\Weak(p+\ktilde,m)~.
\end{eqnarray}
We now substitute the weak field expansion of the Schwinger propagator from Eq.~\eqref{eq.Schwinger.Weak} in the above equation to get,
\begin{eqnarray}
\mcV_\Weak(k,p) &=& \mcV_\vac(k,p) - (eB)^2ig^3\!\!\int\!\!\frac{d^4\ktilde}{(2\pi)^4}\frac{1}{(\ktilde^2-M^2+i\epsilon)}
\TB{\frac{(\pper+\ktilper)^2-(\ppll+\ktilpll)^2+m^2}{\{(k-\ktilde)^2-m^2+i\epsilon\}\{(p+\ktilde)^2-m^2+i\epsilon\}^4} \right. \nn \\
&& \left. \hspace{7cm}
\frac{(\kper-\ktilper)^2-(\kpll-\ktilpll)^2+m^2}{\{(k-\ktilde)^2-m^2+i\epsilon\}^4\{(p+\ktilde)^2-m^2+i\epsilon\}}}~.
\end{eqnarray}
Using standard Feynman parametrization, we combine the denominators of the above equation and get,
\begin{eqnarray}
\mcV_\Weak(k,p) &=& \mcV_\vac(k,p) -(eB)^220ig^3 \!\!\int\!\!\frac{d^4\ktilde}{(2\pi)^4}\int_{0}^{1}\!\!\int_{0}^{1}\!\!\int_{0}^{1}\!\!dxdydz\delta(1-x-y-z)z^3
\nn \\ && \hspace{4cm} \times \TB{
\frac{(\pper+\ktilper)^2-(\ppll+\ktilpll)^2+(\kper-\ktilper)^2-(\kpll-\ktilpll)^2+2m^2}{[(\ktilde-yk+zp)^2-\Delta_\mcV]^6}}
\end{eqnarray}
where $\Delta_\mcV$ is defined in Eq.~\eqref{eq.Delta.V.Vac}.
Shifting $\ktilde\rightarrow(\ktilde+yk-zp)$, the momentum integral in the 
above equation could be performed using the identities provided in Appendix~\ref{appendix.identities}. After some simplifications, we arrive at
\begin{eqnarray}
\mcV_\Weak(k,p) &=& \mcV_\vac(k,p) + g^3\frac{(eB)^2}{8\pi^2}\int_{0}^{1}\!\!dy\int_{0}^{1-y}\!\!dzz^3
\TB{\frac{}{}(\pper^2-\ppll^2)(1+z^2)+(\kper^2-\kpll^2)y^2 \right. \nn \\ && \left. \hspace{6cm}
	+2(\kpll\cdot\ppll-\kper\cdot\pper)yz+m^2\frac{}{}}\frac{1}{\Delta_\mcV^4}~.
\end{eqnarray}
It is to be noted that, the $\Delta_\mcV$ in the above equation contains a small negative imaginary part, which will give rise to non-zero 
imaginary part of the vertex function in certain kinematic domains. 
In order to calculate the imaginary part of the vertex function, we use the trick as given in Eq.~\eqref{eq.Trick} and write,
\begin{eqnarray}
\IM\mcV_\Weak(k,p) &=& \IM\mcV_\vac(k,p) - g^3\frac{(eB)^2}{48\pi}\frac{\del^3}{\del\lambda^3}\int_{0}^{1}\!\!dy\int_{0}^{1-y}\!\!dzz^3
\TB{\frac{}{}(\pper^2-\ppll^2)(1+z^2)+(\kper^2-\kpll^2)y^2 \right. \nn \\ && \left. \hspace{0cm}
	+2(\kpll\cdot\ppll-\kper\cdot\pper)yz+m^2\frac{}{}}
\delta\TB{(y+z)^2m^2+(1-y-z)M^2-yz(k+p)^2+\lambda\frac{}{}}\Bigg|_{\lambda=0}~.
\label{eq.Vertex.Weak.1}
\end{eqnarray}
In order to simplify, we transform the Dirac delta function in the above equation as
\begin{eqnarray}
\delta\TB{(y+z)^2m^2+(1-y-z)M^2-yz(k+p)^2+\lambda} = \frac{\delta(z-\ztil_+)+\delta(z-\ztil_-)}{\sqrt{(M^2+q^2y)^2-4m^2(M^2+q^2y^2+\lambda)}}
\label{eq.delta.3}
\end{eqnarray} 
where $q=(k+p)$ and 
\begin{eqnarray}
\ztil_\pm(q^2,y,\lambda) = \frac{1}{2m^2}\TB{M^2-2m^2y+q^2y\pm\sqrt{(M^2+q^2y)^2-4m^2(M^2+q^2y^2+\lambda)}}~. \label{eq.ztilpm.2}
\end{eqnarray}
Substituting Eq.~\eqref{eq.delta.3} into Eq.~\eqref{eq.Vertex.Weak.1} and performing the $dz$ integral using the modified 
Dirac delta function, we get,
\begin{eqnarray}
\IM\mcV_\Weak(k,p) &=& \IM\mcV_\vac(k,p) - g^3\frac{(eB)^2}{48\pi}\sum_{z\in\{\ztil_\pm\}}^{}\frac{\del^3}{\del\lambda^3}
\int_{0}^{1}\!\!dy\Theta(z)\Theta(1-y-z)z^3 \nn \\ && \hspace{2cm}
 \TB{\frac{(\pper^2-\ppll^2)(1+z^2)+(\kper^2-\kpll^2)y^2+2(\kpll\cdot\ppll-\kper\cdot\pper)yz+m^2}
	{\sqrt{(M^2+q^2y)^2-4m^2(M^2+q^2y^2+\lambda)}}}\Bigg|_{\lambda=0}~.
\label{eq.ImVertex.Weak}
\end{eqnarray}
The presence of the step functions in the above equation ensure that the spikes of the Dirac delta functions 
were within the integration domain for a non-vanishing contribution.

\subsection{VERTEX FUNCTION IN THE STRONG FIELD APPROXIMATION} \label{appendix.Vertex.Strong}
In this appendix, we will briefly sketch the calculation of the imaginary part of one-loop $B^0b^+b^-$ vertex function 
under external magnetic field by employing the strong field approximation.
We start with Eq.~\eqref{eq.Vertex.eB}
\begin{eqnarray}
\mcV_\Strong(k,p) &=& ig^3\!\!\int\!\!\frac{d^4\ktilde}{(2\pi)^4}\Delta_F(\ktilde,M)\Delta_\Strong(k-\ktilde,m)\Delta_\Strong(p+\ktilde,m)
\end{eqnarray}
and substitute the strong field Schwinger propagator from Eq.~\eqref{eq.Schwinger.LLL} in the above equation to obtain,
\begin{eqnarray}
\mcV_\Strong(k,p) &=& -4ig^3\!\!\int\!\!\frac{d^4\ktilde}{(2\pi)^4}\exp\TB{\frac{(\kper-\ktilper)^2+(\pper+\ktilper)^2}{eB}} \nn \\ && \hspace{2cm}\times
\TB{\frac{1}{(\ktilde^2-M^2+i\epsilon)\{(\kpll-\ktilpll)^2-m^2-eB+i\epsilon\}\{(\ppll+\ktilpll)^2-m^2-eB+i\epsilon\}}}~.
\end{eqnarray}
Using standard Feynman parametrization, we combine the denominators of the above equation and get,
\begin{eqnarray}
\mcV_\Strong(k,p) &=& -8ig^3\!\!\int\!\!\frac{d^2\ktilper}{(2\pi)^2}\exp\TB{\frac{(\kper-\ktilper)^2+(\pper+\ktilper)^2}{eB}} 
\nn \\ && \hspace{1cm} \times
\int_{0}^{1}\!\!\int_{0}^{1}\!\!\int_{0}^{1}\!\!dxdydz\delta(1-x-y-z)\!\! \int\!\!\frac{d^2\ktilpll}{(2\pi)^2}
\frac{1}{[(\ktilpll-y\kpll+z\ppll)^2-\tilde{\Delta}_{\mcV}]^3}
\end{eqnarray}
where,
\begin{eqnarray}
\tilde{\Delta}_{\mcV} = (y^2-y)\kpll^2+(z^2-z)\ppll^2-2yz(\kpll\cdot\ppll)+(y+z)(m^2+eB)+(1-y-z)(M^2-\ktilper^2)-i\epsilon~.
\end{eqnarray}
Shifting $\ktilpll\rightarrow(\ktilpll+y\kpll-z\ppll)$, the $d^2\ktilpll$ integral in the 
above equation could be performed using the identities provided in Appendix~\ref{appendix.identities}. After some simplifications, we arrive at
\begin{eqnarray}
\mcV_\Strong(k,p) &=& \frac{-g^3}{\pi}\int_{0}^{1}\!\!dy\int_{0}^{1-y}\!\!dz
\int\!\!\frac{d^2\ktilper}{(2\pi)^2}\exp\TB{\frac{(\kper-\ktilper)^2+(\pper+\ktilper)^2}{eB}}\frac{1}{\tilde{\Delta}_\mcV^2} ~.
\end{eqnarray}
Next we go to the cylindrical polar coordinate and make the substitution $\xi=-2\kper^2/eB$ which implies 
$d^2\ktilper=\frac{eB}{4}\kperv d\kperv d\phi$. The integration over the azimuthal angle $d\phi$ could then be analytically performed 
and be expressed in terms of the modified Bessel function of first kind $I_0$ (see Appendix~\ref{appendix.identities}). We finally get
\begin{eqnarray}
\mcV_\Strong(k,p) &=& -g^3\frac{eB}{8\pi^2}\exp\TB{\frac{\kper^2+\pper^2}{eB}}
\int_{0}^{\infty}\!\!d\xi e^{-\xi}I_0\FB{\sqrt{\frac{\xi}{2eB}\SB{\frac{}{}(\kper+\pper)^2-2\kper^2-2\pper^2}}} \nn \\ && \hspace{5cm} \times
\int_{0}^{1}\!\!dy\int_{0}^{1-y}\!\!dz\frac{1}{\tilde{\Delta}_\mcV^2}~.
\end{eqnarray}
It is to be noted that, the $\tilde{\Delta}_\mcV$ in the above equation contains a small negative imaginary part, 
which will give rise to non-zero imaginary part of the vertex function in certain kinematic domains. 
In order to calculate the imaginary part of the vertex function, we use the trick as given in Eq.~\eqref{eq.Trick} and write,
\begin{eqnarray}
\IM\mcV_\Strong(k,p) &=& g^3\frac{eB}{8\pi}\exp\TB{\frac{\kper^2+\pper^2}{eB}}
\int_{0}^{\infty}\!\!d\xi e^{-\xi}I_0\FB{\sqrt{\frac{\xi}{2eB}\SB{\frac{}{}(\kper+\pper)^2-2\kper^2-2\pper^2}}} 
\frac{\del}{\del\lambda}\int_{0}^{1}\!\!dy\int_{0}^{1-y}\!\!dz \nn \\ && \hspace{-1cm} \times
\delta\TB{(y^2-y)\kpll^2+(z^2-z)\ppll^2-2yz(\kpll\cdot\ppll)+(y+z)(m^2+eB)+(1-y-z)(M^2+eB\xi/2)+\lambda}\Bigg|_{\lambda=0}~.
\label{eq.l2}
\end{eqnarray}
In order to simplify, we transform the Dirac delta function in the above equation as
\begin{eqnarray}
&& \delta\TB{(y^2-y)\kpll^2+(z^2-z)\ppll^2-2yz(\kpll\cdot\ppll) + (y+z)(m^2+eB)+(1-y-z)(M^2+eB\xi/2)+\lambda} \nn \\ && \hspace{2cm}
= \frac{1}{\sqrt{\mathcal{D}}}\TB{\delta(z-\Ztil_+)+\delta(z-\Ztil_-)}
\label{eq.delta.4}
\end{eqnarray} 
where 
\begin{eqnarray}
\Ztil_\pm = \frac{1}{4 \ppll^2}\TB{eB \xi -2 eB+4y\kpll\cdot\ppll-2 m^2+2 M^2+2\ppll^2 \pm 2 \sqrt{\mathcal{D}}}
\label{eq.Zpm}
\end{eqnarray}
with 
\begin{eqnarray}
\mathcal{D} &=& \TB{\frac{1}{2} eB (\xi -2)+2y\kpll\cdot\ppll -m^2+M^2+\ppll^2}^2 \nn \\ &&
-4 \ppll^2 \TB{\frac{1}{2} eB \{\xi (1-y) +2 y\} +\kpll^2 y^2-\kpll^2 y+m^2 y-M^2 (y-1)+\lambda}~.
\end{eqnarray}
Substituting Eq.~\eqref{eq.delta.4} into Eq.~\eqref{eq.l2} and performing the $dz$ integral using the modified 
Dirac delta function, we get,
\begin{eqnarray}
\IM\mcV_\Strong(k,p) &=& g^3\frac{eB}{8\pi}\exp\TB{\frac{\kper^2+\pper^2}{eB}}
\int_{0}^{\infty}\!\!d\xi e^{-\xi}I_0\FB{\sqrt{\frac{\xi}{2eB}\SB{\frac{}{}(\kper+\pper)^2-2\kper^2-2\pper^2}}} 
\nn \\ && \hspace{1cm} \times
\frac{\del}{\del\lambda}\int_{0}^{1}\!\!dy
\frac{\Theta(\Ztil_+)\Theta(1-y-\Ztil_+)+\Theta(\Ztil_-)\Theta(1-y-\Ztil_-)}{\sqrt{\mathcal{D}}}
\Bigg|_{\lambda=0}~.
 \label{eq.ImVertex.Strong}
\end{eqnarray}
The presence of the step functions in the above equation ensure that the spikes of the Dirac delta functions 
were within the integration domain for a non-vanishing contribution. 
\bibliographystyle{apsrev4-1}
\bibliography{snigdha}

\end{document}